\documentclass[aps,pra,epsf,superscriptaddress,amsmath,amssymb,amsfonts,twocolumn,showpacs,floatfix]{revtex4-2}

\usepackage[utf8]{inputenc}

\usepackage{amsmath}
\usepackage{amssymb}
\usepackage{amsfonts}
\usepackage{amstext}
\usepackage{amsthm}
\usepackage{mathtools}
\usepackage{physics}

\usepackage{bm} 

\usepackage{graphicx}
\graphicspath{{figures/}}

\usepackage{epsfig}
\usepackage{dcolumn}
\usepackage{bm}
\usepackage{braket}
\usepackage{color}
\usepackage[colorlinks=true,citecolor=blue]{hyperref}

\usepackage[english]{babel}

\usepackage{multirow}

\usepackage[dvipsnames]{xcolor}

\definecolor{deeppurple}{rgb}{0.7, 0, 0.8}

\allowdisplaybreaks
\advance\parskip 0.4pt

\usepackage{hyperref}
\hypersetup{
	urlcolor=blue
}

\usepackage{natbib}

\newcommand{\be}{\begin{equation}}
\newcommand{\ee}{\end{equation}}

\begin{document}

\title{Existence, stability and spatio-temporal dynamics of time-quasiperiodic solutions
 on a finite background in discrete nonlinear Schr\"odinger models}

\author{E.~G. Charalampidis}
\affiliation{Mathematics Department,
California Polytechnic State University,
San Luis Obispo, CA 93407-0403, USA}

\author{G. James}
\affiliation{
Univ. Grenoble Alpes, CNRS, Inria, Grenoble INP * , LJK, 38000 Grenoble, France * Institute of Engineering Univ. Grenoble Alpes
}

\author{J.~Cuevas-Maraver}
\affiliation{Grupo de F\'{i}sica No Lineal, Departamento de F\'{i}sica Aplicada I,
Universidad de Sevilla. Escuela Polit\'{e}cnica Superior, C/ Virgen de \'{A}frica,
7, 41011-Sevilla, Spain\\
Instituto de Matem\'{a}ticas de la Universidad de Sevilla (IMUS).
Edificio Celestino Mutis. Avda. Reina Mercedes s/n, 41012-Sevilla, Spain}

\author{D. Hennig}
\affiliation{Department of Mathematics, University of Thessaly, Lamia 35100, Greece}

\author{N.~I. Karachalios}
\affiliation{Department of Mathematics, University of Thessaly, Lamia 35100, Greece}

\author{P.~G. Kevrekidis}
\affiliation{Department of Mathematics and
Statistics, University of Massachusetts
Amherst, Amherst, MA 01003-4515, USA}


\begin{abstract}
In the present work we explore the potential of models of the discrete
nonlinear Schr{\"o}dinger (DNLS) type to support spatially localized and
temporally quasiperiodic solutions on top of a finite background.
Such solutions are rigorously shown to exist in the vicinity of the anti-continuum,
vanishing coupling limit of the model. We then use numerical continuation to illustrate
their persistence for finite coupling, as well as to explore their spectral stability.
We obtain an intricate bifurcation diagram showing a progression of such
solutions from simpler ones bearing single- and two-site excitations to more
complex, multi-site ones with a direct connection of the branches of the self-focusing
and self-defocusing nonlinear regime. We further probe the variation
of the
solutions  obtained towards the limit of vanishing frequency for both signs
of the nonlinearity. Our analysis is complemented by exploring the dynamics of the solutions
via direct numerical simulations.
\end{abstract}

\maketitle

%

\section{Introduction}
The topic of nonlinear dynamical lattices and their localized
modes has received considerable attention over the past 4 decades,
especially since the illustration of their generic emergence in
anharmonic crystals~\cite{takeno}, and also the rigorous mathematical
proof of their existence under suitable non-resonance conditions~\cite{mackay}.
Indeed, relevant progress has been by now summarized in a number of influential
reviews such as~\cite{Flach2008,Aubry2006}. Importantly, beyond the mathematical
and computational analyses of the existence of such modes, a key reason for their
extensive study has, arguably, been the impact of associated prototypical models
in advancing our understanding as concerns light propagation in optical waveguides~\cite{LEDERER20081}
and mean-field models of atomic condensates in optical lattices~\cite{RevModPhys.78.179}.
In both of these central applications, a prototypical dynamical model that has
contributed to the analysis, simulations and experimental progress
has been the discrete nonlinear Schr{\"o}dinger (DNLS)
equation~\cite{kev09}.

On the other hand, a topic of ever growing interest over the last
(especially) 15 years has been the dynamics of rogue or freak
waves~\cite{Kharif2009}. While observations of such have existed for
over half a century~\cite{Draper1966} and the well-known measurement
of the Draupner wave on the first day of 1995 has captured the attention
of the physics, engineering and mathematics communities alike, it has been
mostly over the last decade and a half that numerous relevant developments
have arisen especially on the nonlinear analysis of such waves, as motivated
by carefully-controlled experiments. Indeed, the leveraging of novel detection
techniques to observe them in optical systems~\cite{Solli2007,Solli2008} has led
to numerous further explorations within that field~\cite{Kibler2010,Kibler2012,DeVore2013,Frisquet2016}.
In parallel, the ability to produce fluid experiments of fundamental
and higher-order rogue waves emerged in the works of~\cite{Chabchoub2011,Chabchoub2012,Chabchoub2014} and led to
the re-creation of the Draupner wave in~\cite{ton_2019}. In addition
to the broader relevance of these ideas as argued, e.g., in other fields such
as plasmas~\cite{Bailung2011}, and very recently
superfluids~\cite{romeroros2023experimental}, the maturation of these efforts
can be recognized in a number of impactful reviews
such as~\cite{Yan2012a,Onorato2013,Dudley2014,Mihalache2017,natrevphys}.

Our original aim in the present work was to explore the potential
inter-connection between these important current research themes.
Indeed, this has been an ongoing effort
that has identified
analogues of rogue-type structures
(such as the famous Peregrine (P) soliton~\cite{Peregrine1983}, the
Kuznetsov-Ma (KM) soliton~\cite{Kuznetsov1977,Ma1979} or the Akhmediev
breather (AB)~\cite{Akhmediev1986}) in the integrable discrete realm of
the so-called Ablowitz-Ladik (AL) model~\cite{sotoc}. More recent
efforts
from a subset of the present authors
have attempted to leverage the so-called Salerno model~\cite{salerno1992quantum}
to homotopically interpolate
between the AL the physically realistic DNLS
model~\cite{HOFFMANN20183064,sullivan}. However, the relevant computational continuation
efforts at the bifurcation level typically encountered turning points, leading them to discover
unprecedented AL solutions~\cite{sullivan} bearing
an oscillatory background, without necessarily improving our
understanding of the DNLS (or the continuum, for that matter) limit.
Motivated by this finding, we raise the question of whether such
periodic solutions can be identified in DNLS-type models that sit on top of
a finite, i.e., flat background. It is important to highlight here
that leveraging the phase invariance of the DNLS model and factoring
out
a constant background, one can seek time-periodic waveforms which
are {\it quasiperiodic} ones in the original model, in a way
reminiscent of the works of~\cite{Johansson_1997}
and~\cite{pgkmiw} which, however,
sought such solutions on top of a vanishing background or of a stationary soliton, respectively.

Motivated by the above observations, we start from the highly
controllable anti-continuum (AC) limit of vanishing coupling across the
lattice nodes. We show that in the neighborhood of such a limit,
breathing-in-time solutions (in the frame ``co-rotating'' with a certain
frequency, hence quasiperiodic in the original frame)
can be {\it rigorously} shown to exist.
We then corroborate these findings through numerical computations that
reveal an intricate bifurcation structure connecting such breathing states
between the focusing and the defocusing DNLS settings. Although our dynamical
simulations reveal the nature of the solutions we identified, it is
important to distinguish
their characteristics from  the inherent features of rogue waves.
The latter include high-amplitude waveforms that ``appear out of nowhere and disappear
without a trace'', while exceeding a
certain amplitude threshold~\cite{enwiki:1154847066}. Instead, our
waveforms will be weakly quasiperiodically breathing on top of a
fixed density profile.
Regardless, the obtained waveforms, while only motivated by the
rogue patterns, constitute a novel class of quasiperiodic solutions of the
experimentally relevant DNLS equation and as such are of potential
interest, including in experiments within nonlinear optics (optical
waveguides)~\cite{LEDERER20081} and atomic physics
(Bose-Einstein condensates in optical lattices)~\cite{RevModPhys.78.179}.
Our presentation is structured as follows. In section II, we present
the rigorous proof of existence of the states of interest, while in
section III, we detail our numerical continuation, spectral stability
and nonlinear dynamics results. Finally, in section IV, we summarize
our findings and present our conclusions.

\section{Rigorous Analysis of time-quasiperiodic solutions to DNLS models}
We consider the general model
\begin{equation}
\label{eq:gdnls}
i\, \dot{\psi}_n + \psi_n\, f(|\psi_n|^2) + \sum_{p\in \mathbb{Z}}{K_{n-p}\, \psi_p}=0,
\quad n\in \mathbb{Z} ,
\end{equation}
where $f \in C^1 ((0,\infty),\mathbb{R})$. We denote by
$\ell_1(\mathbb{Z},\mathbb{K})$ the classical Banach space of summable
sequences in $\mathbb{K}=\mathbb{R}$ or $\mathbb{C}$.
We assume that the sequence $K=(K_n)_{n\in \mathbb{Z}} \in \ell_1(\mathbb{Z},\mathbb{R})$
satisfies $\sum_{n\in\mathbb{Z}}{K_n}=0$,
and denote the closed linear subspace of $\ell_1(\mathbb{Z},\mathbb{R})$
consisting of zero-sum sequences by $S_0$.
This class of models encompasses
the generalized form of the DNLS equation~\cite{kev09} corresponding to
$K_0=-2d$, $K_{\pm 1}=d$ and $K_n=0$ elsewhere, with
$d$ being a coupling parameter.

With the zero-sum assumption on the sequence $K$, Eq.~\eqref{eq:gdnls} admits
time-periodic solutions of the form
\begin{equation}
\label{syncosc}
\psi_n (t) =  R\, e^{i\, (\Omega\, t + \varphi)},
\end{equation}
with nonvanishing amplitude $R>0$, frequency $\Omega = f(R^2 )$ and phase shift $\varphi$.
In what follows, we consider a solution of the form~\eqref{syncosc} with $\varphi=0$ and
assume that the nondegeneracy condition $f^\prime (R^2)\neq 0$ is satisfied. This holds
still, in particular, for the classical cubic nonlinearities, where $f^\prime$ is a nonvanishing
constant.

We look for breather solutions of Eq.~\eqref{eq:gdnls} on a background, corresponding to
spatially localized perturbations of solution~\eqref{syncosc}, quasiperiodic in time with two
fundamental frequencies. For this purpose, we set
\begin{eqnarray}
  \psi_n (t)= e^{i\, \Omega\, t }\, u_n(t),
  \label{eq0}
\end{eqnarray}
with $u_n$ being time-periodic with frequency $\omega_b = f(A^2)-\Omega$, and for some fixed
$A>0$. Substitution in Eq.~\eqref{eq:gdnls} yields
\begin{equation}
\label{eq:gdnlsu}
i\, \dot{u}_n + u_n\, [\, f(|u_n|^2)-\Omega \, ]+ \sum_{p\in \mathbb{Z}}{K_{n-p}\, u_p}=0,
\quad n\in \mathbb{Z} .
\end{equation}
We assume $f^\prime (A^2)\neq 0$, and the nonresonance condition $f(A^2) \neq \Omega$, i.e.,
$\omega_b \neq 0$ (so that there are two distinct frequencies in our solution).
In addition, we consider time-reversible solutions satisfying $u_n (-t)=\bar{u}_n(t)$, where
the bar notation is used for complex conjugate.

In the anti-continuum (AC) limit $K=0$~\cite{mackay}, the system given by Eq.~\eqref{eq:gdnlsu}
becomes uncoupled, and admits solutions $u_n^0$ taking the form
\begin{equation}
\label{sol:uncpl}
u_n^0 (t) =
\left\{
\begin{array}{ll}
A\, e^{i\, \omega_b t } &\mbox{if~} n\in \mathbb{I}^+,\\
-A\, e^{i\, \omega_b t } &\mbox{if~} n\in \mathbb{I}^-,\\
R& \mbox{if~} n\in \mathbb{I},
\end{array}
\right.
\end{equation}
where $\mathbb{I}^+$, $\mathbb{I}^-$ are arbitrary finite subsets of
$\mathbb{Z}=\mathbb{I}\cup \mathbb{I}^+\cup \mathbb{I}^-$.
In the sequel, we search for solutions $u= (u_n(.))_{n\in \mathbb{Z}}$
of Eq.~\eqref{eq:gdnlsu} close to the form~\eqref{sol:uncpl} when $K$ is small
in $\ell_1(\mathbb{Z})$. For this purpose, we
use the ansatz
\begin{equation}
\label{sol:perturbed}
u_n = u_n^0 + y_n,
\end{equation}
where the perturbation
$y= (y_n(.))_{n\in \mathbb{Z}}$
is sought in a spatially localized and
time-periodic form with period $T=2\pi / \omega_b$.
More precisely, we define for all $m\geq 0$ the Banach space
$$
X_m = \left\{ \,
y \in C^m \big( \mathbb{R}/T\mathbb{Z}, \ell_1(\mathbb{Z},\mathbb{C}) \big), \,
y_n (-t)=\bar{y}_n(t)
\, \right\}
$$
($X_m$ is endowed with the usual $C^m$ norm)
and we assume $y \in X_1$.
We substitute expression~\eqref{sol:perturbed}
in Eq.~\eqref{eq:gdnlsu} and obtain
the following equation for the perturbation $y$:
\begin{eqnarray}
\nonumber
0&=&
i\, \dot{y}_n
+ u_n^0\, \big(\, f(|u_n^0+y_n|^2)-f(|u_n^0|^2 \, \big) \\
\label{eq:gdnlsy}
&&+ y_n\, \big(\, f(|u_n^0+y_n|^2)-\Omega \, \big)\\
\nonumber
&&
+ \sum_{p\in \mathbb{Z}}{K_{n-p}\, (u_p^0+y_p)},
\quad n\in \mathbb{Z} .
\end{eqnarray}
Let us identify $y$ with $\tilde{y}=(\operatorname{Re}(y),\operatorname{Im}(y))\in \tilde{X}_1$,
where for all $m\geq 0$
\begin{eqnarray*}
\tilde{X}_m &=& \left\{ \,
(a_n(.),b_n(.))_{n\in \mathbb{Z}} \in C^m \big( \mathbb{R}/T\mathbb{Z}, \ell_1(\mathbb{Z},\mathbb{R})^2 \big), \right. \\
& & \left. a_n (-t)={a}_n(t), \ b_n (-t)=-{b}_n(t)
\, \right\} .
\end{eqnarray*}
System~\eqref{eq:gdnlsy} can be considered as a nonlinear equation
$F(\tilde{y},K)=0$ with $F \in C^1(\tilde{X}_1 \times S_0,\tilde{X}_0 )$
defined by the right-hand-side of Eq.~\eqref{eq:gdnlsy} with $F(0,0)=0$.

In order to solve Eq.~\eqref{eq:gdnlsy} for $K \approx 0$ using the implicit function theorem,
we need to check the invertibility of $L=D_{\tilde{y}}F(0,0) \in \mathcal{L}(\tilde{X}_1,\tilde{X}_0)$.
Let $\tilde{f} = (a_n(.),b_n(.))_{n\in \mathbb{Z}} \in \tilde{X}_0$
and search for $\tilde{y} \in \tilde{X}_1$ such that $L\, \tilde{y}=\tilde{f}$.
This problem can be rewritten as
\begin{equation}
\label{eq:lin}
f_n =
\left\{
\begin{array}{l}
i\, \dot{y}_n + R^2 f^\prime (R^2)\, (y_n+\bar{y}_n), \quad n\in \mathbb{I},\\
i\, \dot{y}_n + A^2 f^\prime (A^2)\, (y_n+e^{2i\, \omega_b t }\bar{y}_n)+\omega_b  y_n,\
n\in \mathbb{Z}\setminus \mathbb{I},
\end{array}
\right.
\end{equation}
with $f_n = a_n +i\, b_n$,  $f_n (-t)=\bar{f}_n(t)$.
We start by solving the above equations for any given $n\in \mathbb{I}$.
Expanding $y_n$, $f_n$ in Fourier series
(omitting the index $n$ in the Fourier coefficients for notational simplicity), we have
$$
y_n(t) = \sum_{k\in\mathbb{Z}}{c_k\, e^{i k \omega_b t }},
\quad
f_n(t) = \sum_{k\in\mathbb{Z}}{b_k\, e^{i k \omega_b t }},
$$
with $c_k , b_k \in \mathbb{R}$ due to time-reversibility symmetry.
Substitution of the above expansions in Eq.~\eqref{eq:lin} yields
\begin{subequations}
\begin{align}
c_0 &= b_0 / (2R^2 f^\prime (R^2)), \\
c_k &= \frac{R^2 f^\prime (R^2)}{k^2 \omega_b^2}(b_{-k} - b_{k})-\frac{b_k}{k \omega_b},
\end{align}
\end{subequations}
for all $k\neq 0$. Similarly, for $n\in \mathbb{Z}\setminus \mathbb{I}$ we obtain
\begin{subequations}
\begin{align}
c_1 &= b_1 / (2A^2 f^\prime (A^2)), \\
c_k &= \frac{A^2 f^\prime (A^2)}{(k-1)^2 \omega_b^2}(b_{2-k} - b_{k})-\frac{b_k}{(k-1) \omega_b},
\end{align}
\end{subequations}
for all $k\neq 1$. This yields a unique solution $y\in X_1$ to Eq.~\eqref{eq:lin}, where
$C^1$-regularity follows from a standard bootstrap argument. As a result, the linearized
map $L$ is invertible.

Consequently, by the implicit function theorem, the solution $y=0$ to Eq.~\eqref{eq:gdnlsy}
with $K = 0$ can be continued for $K \approx 0$ into a unique solution $y=Y(K) \in X_1$, where
$Y$ is a $C^1$ map defined on a neighborhood of $K=0$ in $S_0$ and $Y(0)=0$. Equivalently, for
all $K \approx 0$ in $S_0$, Eq.~\eqref{eq:gdnlsu} admits a unique $T$-periodic reversible solution
such that $\| u - u^0 \|_{X_1}$ is small, which depends smoothly on $K$.

We conclude this analysis 
by briefly mentioning some symmetry considerations when $K_{-n}=K_n$
(as in the generalized DNLS equation), a case when Eq.~\eqref{eq:gdnlsu} has
the invariance $n \rightarrow -n$. If the solution $u^0$ in the AC limit
is site-centered, i.e. $u^0_{-n}(t)=u^0_n(t)$, then by uniqueness of the local continuation
$u$ one has also $u_{-n}(t)=u_n(t)$. This case occurs in particular for
$u_0^0 (t) = \pm A\, e^{i \omega_b t }$ and $u_n^0 (t) = R$ elsewhere.
Similarly, if $u_0$ is bond-centered, i.e. $u^0_{n}(t)=u^0_{1-n}(t)$, then
the local continuation $u_n (t)$ has the same symmetry. This is the case in
particular if $u_0^0 (t) =u_1^0 (t) = \pm A\, e^{i \omega_b t }$ and $u_n^0 (t) = R$
elsewhere.

\section{Numerical Computations}
While our analysis has been kept quite general to illustrate the
breadth of the relevant ideas, in the numerical computations
that follow, motivated by the experimental realizability
of the nearest-neighbor, cubic DNLS model~\cite{LEDERER20081,RevModPhys.78.179},
we restrict our considerations to the latter, i.e., $f(|u_n|^2)=|u_n|^2$
and $K_{n-p}=d \delta_{n-p,\pm 1}- 2d \delta_{n-p,0}$ (where the Kronecker-$\delta$
is implied). This way, Eq.~\eqref{eq:gdnlsu} reduces to:
\begin{align}
\label{dnls_numer_1}
i\dot{u}_{n}+d\left(u_{n+1}-2u_{n}+u_{n-1}\right)+\left(|u_{n}|^{2}-\Omega\right)u_{n}=0.
\end{align}
In the numerical computations discussed in this section, we identify time-periodic
solutions $u_{n}(t)=u_{n}(t+T)$ with period $T=2\pi/\omega_{b}$ (or, equivalently with
frequency $\omega_{b}$)
to Eq.~\eqref{dnls_numer_1}. We carry out our computations on a lattice with $N=50$
sites where periodic boundary conditions are imposed, i.e., $u_{-N/2}=u_{N/2}$.
In our visualization of the relevant waveform, we normalize its
background
density (i.e., square modulus equal to $\Omega$ which is set to $1$ hereafter,
i.e., $\Omega\equiv 1$). Similarly to~\cite{sullivan}, time-periodic
solutions
(on top of the background frequency $\Omega$) are sought by
using the ansatz:
\begin{align}
\label{ansatz_Fourier_in_time}
u_{n}(t)=1+ \sum_{k=-\infty}^{\infty} {\cal U}_{n,k}e^{i k \omega_{b}t}.
\end{align}
Upon substituting this
to the DNLS Eq.~\eqref{dnls_numer_1}
leads to a root-finding problem for the Fourier coefficients ${\cal U}_{n,k}$ that is solved by
means of Newton's method. We note that we truncated the series [cf. Eq.~\eqref{ansatz_Fourier_in_time}]
by fixing $|k|\leq m$ with $m=21$,
hence considering $2m+1=43$ Fourier modes in time. Those were proven to be enough to
resolve the $1+0i$ Floquet multiplier associated with the
linearization
of the DNLS around the solution, a mode that is theoretically expected
to
be present due to the Hamiltonian nature of the model. Upon convergence in Newton's
method, we reconstruct the solution $u_{n}$ at $t=0$ by summing over the Fourier modes according to
Eq.~\eqref{ansatz_Fourier_in_time}, and then using the resulting
$u_n(t)$
in Eq.~(\ref{eq0}).
We explore the configuration space of time-periodic solutions to the DNLS equation by performing a pseudo-arclength
continuation~\cite{kuznetsov_book_2023} over the coupling parameter $d$ and frequency $\omega_{b}$. At each
continuation step, a Floquet stability analysis is carried out through
the solution of the variational equations for the associated
monodromy matrix, in order to determine the stability characteristics
of the solutions we found; see~\cite{sullivan} for further details about the setup of the stability problem.

In Fig.~\ref{fig1}, we summarize our numerical results for site- and bond-centered time-periodic
solutions that our solvers converged to. In particular, we show branches of time-periodic solutions
with $\omega_{b}=8$ and background amplitude $\Omega=1$
where the set $\mathbb{I}^-$ consists of a single (top panel) and double
(bottom panel) site near the AC limit
(and $\mathbb{I}^+$ is empty),
analogously to the well-known
site-centered
and bond-centered solutions (on top of a vanishing background)
of the DNLS model~\cite{kev09}. As a relevant bifurcation diagnostic,
we measure the average norm of the
solutions $\langle{N\rangle}=\sum_n\sum_{k=-m}^{m}|{\cal U}_{n,k}|^2$, and depict the densities
$|u_{n}|^{2}$ of the solutions' spatial profile (through its reconstruction from Eq.~\eqref{ansatz_Fourier_in_time})
and their Floquet spectra in Fig.~\ref{fig1}.

Let us begin our discussion by going through the results on site-centered
breathers as they are summarized in the top panel of Fig.~\ref{fig1}. It can be discerned
from the figure that a state of this type at the AC limit can be continued over the coupling strength
$d$, see, Fig.~\ref{fig1}(a), and up to $d\approx 0.4$ before
encountering a turning
point
that leads to a branch
%
with practically a central and a number of lateral excited sites
as this is shown in Fig.~\ref{fig1}(b).
The relevant continuation goes through $d=0$ over
to the defocusing nonlinearity setting of $d<0$ (discussed in more
detail next), and then coming back to the focusing regime again with states
involving more intense lateral excitations as these are progressively
shown in Figs.~\ref{fig1}(c) and (d).
It is interesting to observe that every second one of these
branches features a node that is nearly of vanishing amplitude;
note that a similar feature is present in the panels (e)--(h) of the defocusing
problem below.
All  4 of these sets of panels
are for the same coupling of $d=0.2$. The same sequence
of defocusing and focusing segments continues for multiple
additional turning points (the remaining ones of which
occur around $d \approx 0.35$).

In a similar vein, but now for $d=-0.1$, the top panel of
Fig.~\ref{fig1} also shows four examples of the branches of
(site-centered) time-periodic waveforms for the case of a
defocusing nonlinearity, see the panels (e)-(h) therein.
Indeed, we observe a similar pattern, namely that as one
moves through the different portions of the relevant
branches and the associated turning points (compare the panels
(e) and (f) with (b) and (a), respectively, of the focusing regime).
Gradually more nodes deviate from the background amplitude, forming a
progressively more delocalized breathing excitation as is shown
in panels (g) and (h) therein.
Here, though, contrary to the focusing case where the additional
excitations are higher than the background, the additional
excited sites have densities below those of the background.
Furthermore, it is relevant
to also make some additional observations. Firstly, we remark
that each pair of focusing branches in the diagram segues into
a pair (again involving a turning point) of  defocusing branches,
then on to another focusing loop and so on. Yet, it is interesting
to also examine how the change of stability of the relevant waveforms
occurs along this continuation, as shown in the bottom 
of each panel in Fig.~\ref{fig1}, with the latter representing the
corresponding
Floquet multipliers of the monodromy matrix.

For the focusing portions, the branch is modulationally
unstable, in line with the earlier associated
calculation of~\cite{kivpey} (see also~\cite{sullivan}),
as is reflected in the real Floquet multipliers of
the associated linearization around the periodic orbit;
see Figs.~\ref{fig1}(a)-(d). Yet, as the AC limit of $d=0$
is approached, the modulational instability (MI) band shrinks
and subsequently the relevant multipliers reside on
the unit circle for the defocusing case of $d<0$,
as shown in Figs.~\ref{fig1}(e)-(h). Nevertheless,
it is interesting to highlight that the
(site-centered) breathing waveforms for the case of
of Fig.~\ref{fig1}(e) (and more generally
in the $d<0$ branches) are very weakly unstable
due to {\it isolated} unstable pairs of real
multipliers. The number of pairs increases by one
for every higher branch considered (two in Fig.~\ref{fig1}(f),
three in Fig.~\ref{fig1}(g), four in Fig.~\ref{fig1}(h), etc.
within Fig.~\ref{fig1}).

We briefly highlight the numerical results for bond-centered breathers
as they are shown in the bottom panel of Fig.~\ref{fig1}. Upon analytically
constructing the relevant state from the AC limit, see Fig.~\ref{fig1}(i),
the bifurcation diagram reveals similar features of the states presented herein
with the site-centered ones as regards their background structure and stability
characteristics (compare the site-centered states shown in the top panel of the
figure with the bond-centered ones of the bottom panel, respectively). Also, we
notice that the bond-centered breathers are similarly modulationally unstable in
the focusing regime although they become only weakly unstable in the defocusing regime
similarly to their site-centered counterparts. Again, we highlight that the instability
in the defocusing regime is due to the existence of an isolated unstable pair of real
multipliers (see, Fig.~\ref{fig1}(l)), and the number of the unstable pairs increases
by one as we move to higher branches, see, indicatively, Fig.~\ref{fig1}(m).
It is interesting to observe that the lowest defocusing bond-centered
branch (l) in the bottom panel of Fig.\ref{fig1} has the same number of
unstable eigendirections as the site-centered branch of the defocusing problem
in panel (e) of the top part of Fig.~\ref{fig1}.

Subsequently, we were interested in exploring the
approach of the relevant branches of site-centered solutions
towards the limit where the frequency $\omega_b$ of
the time-periodic solution approaches $0$.
While the patterns obtained are not rogue waves, the latter limiting
process for rogue waves holds particular interest as it turns the
so-called KM solutions into the limiting Peregrine solitonic structure.
To explore this in Fig.~\ref{fig3}, we performed
a continuation both for the focusing case (top panel)
and for the defocusing one (bottom panel) towards the
vanishing frequency limit, indeed for different values
of the coupling constant $d$. Naturally, for none
of the cases, were we able to reach the limit (as
the solution loses its periodic orbit character).
Yet, it was interesting to observe that while the
focusing branches could be continued in a concave
down form towards this limit in the (average) ``power'' dependence
(the sum of the square intensities of the Fourier coefficients) vs.
the frequency $\omega_{b}$, this was not the case in the defocusing setting. In the
latter, a turning point always appeared which is also
tantamount to a stability change, in line with the classic
criterion of~\cite{vakhitov}. The relevant defocusing model
turning point occurred closer to frequencies $\omega_b \rightarrow 0$,
the smaller (in absolute value) the coupling strength $d$ was.

{\protect{\onecolumngrid
\begin{center}
\begin{figure}[!pt]
\begin{center}
\includegraphics[height=.37\textheight, angle =0]{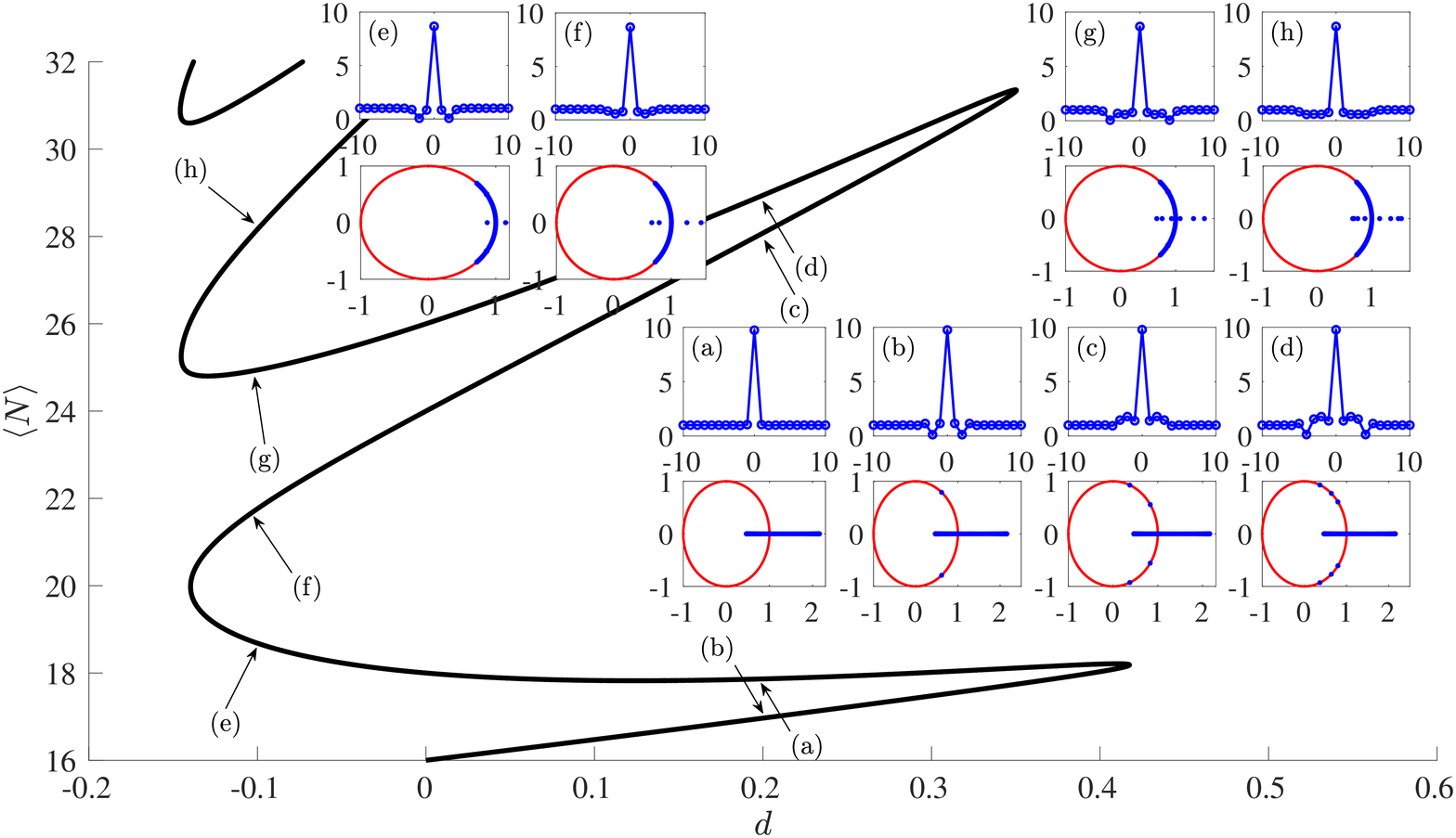}
\includegraphics[height=.37\textheight, angle =0]{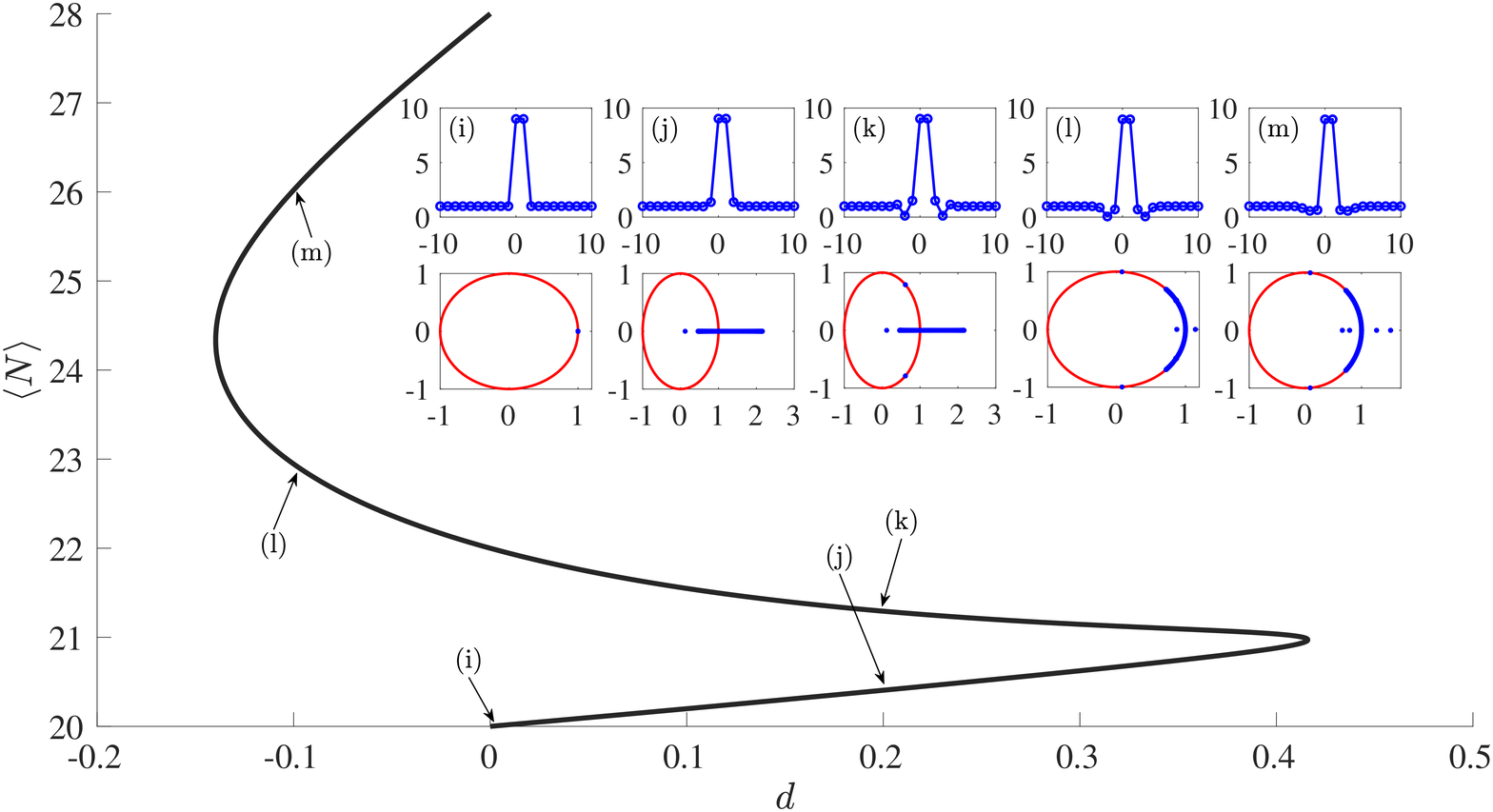}
\end{center}
\caption{(Color online)
Bifurcation diagrams (showcasing the dependence of the average norm of the solutions
$\langle{N\rangle}$ vs the coupling constant $d$) and associated solutions as well as
their Floquet spectra for site- (top panel) and bond-centered (bottom panel) time-periodic
solutions to the DNLS with $w_{b}=8$ and $\Omega=1$. The insets in both panels showcase
the density, i.e., the modulus square $|u_n|^2$ of the solutions found together with their
spectra. The labels therein are associated with the arrows in the respective bifurcation diagrams
(see text for details). For the site-centered breathers, the panels (a)-(d) depict 4 examples
of solutions in the focusing problem whereas the panels (e)-(h) showcase 4 such in the defocusing
problem. In the bond-centered case, in addition to the AC limit profile (i), we demonstrate 2
examples in each of the focusing (see (j) and (k)) and defocusing (see (l) and (m)) regimes.
}
\label{fig1}
\end{figure}
\end{center}
}}
\twocolumngrid

\begin{figure}[!pt]
\begin{center}
\includegraphics[height=.20\textheight, angle =0]{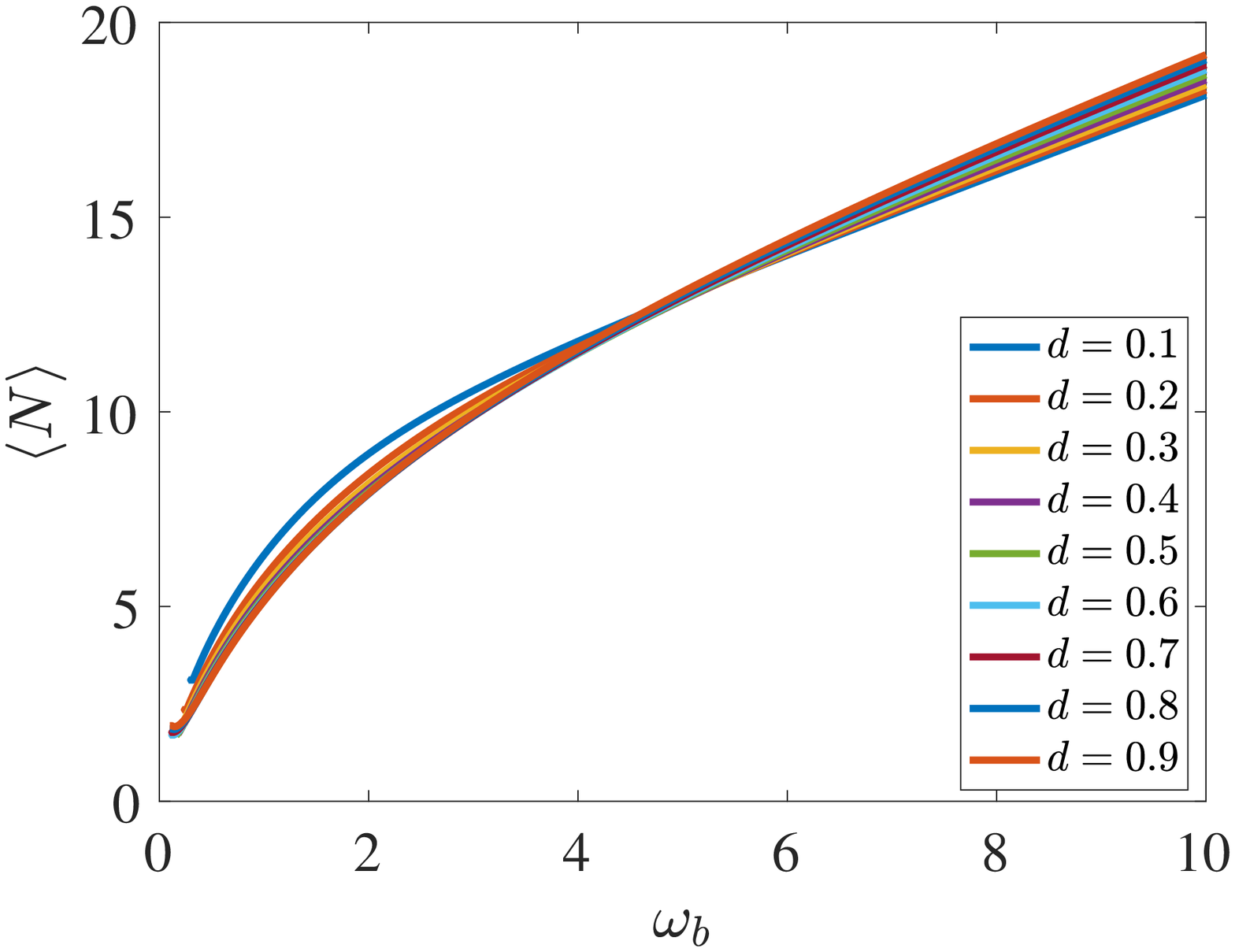}
\includegraphics[height=.20\textheight, angle =0]{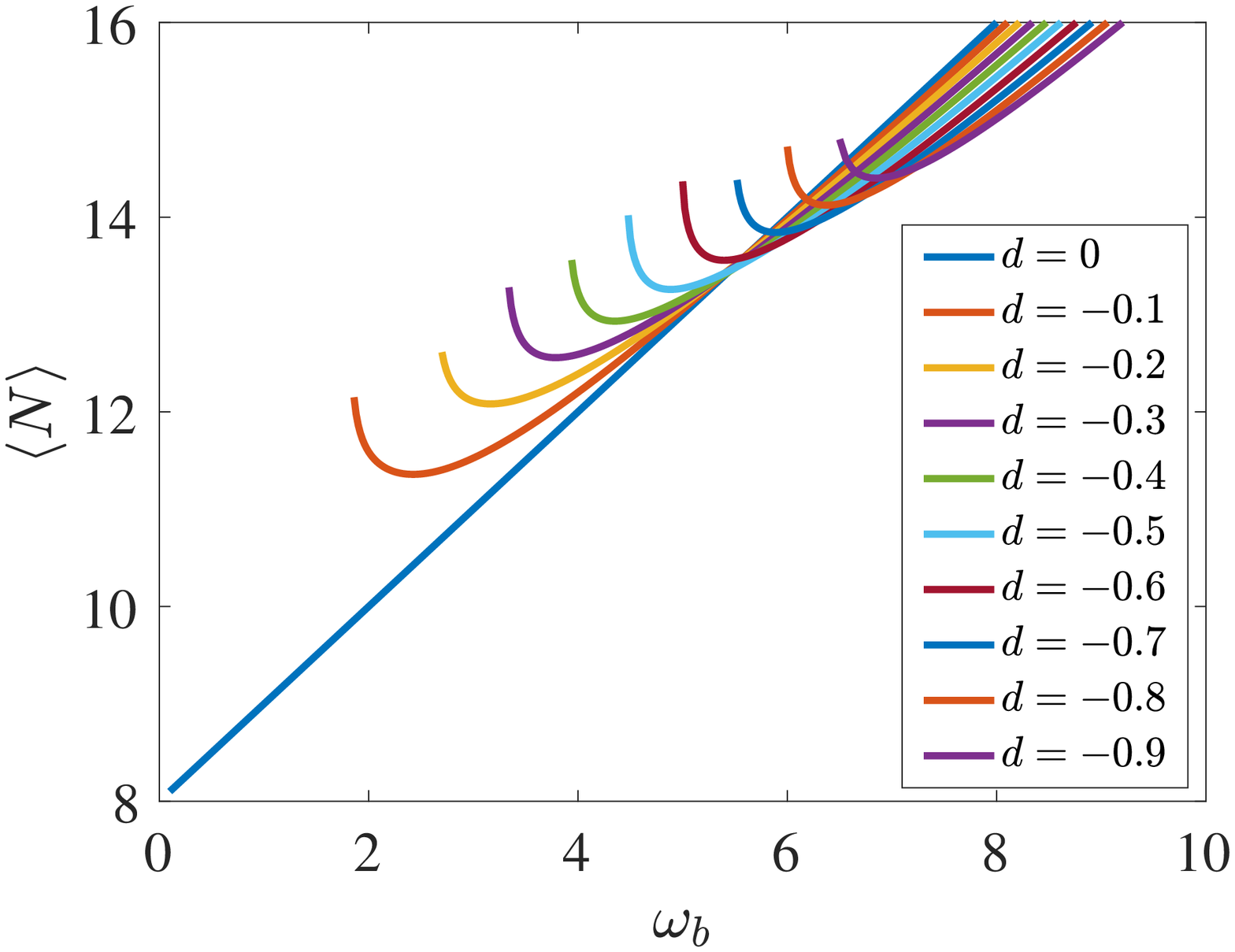}
\end{center}
\caption{(Color online) The average power (sum of the square
intensities of the Fourier coefficients) of the solution as a
function of the frequency $\omega_b$ for the focusing case
(top panel) and the defocusing one (bottom panel). Different
values of the coupling constant $d$ are illustrated by the different
curves, according to the legend.}
\label{fig3}
\end{figure}

\begin{figure}[!pt]
\begin{center}
\includegraphics[height=.135\textheight, angle =0]{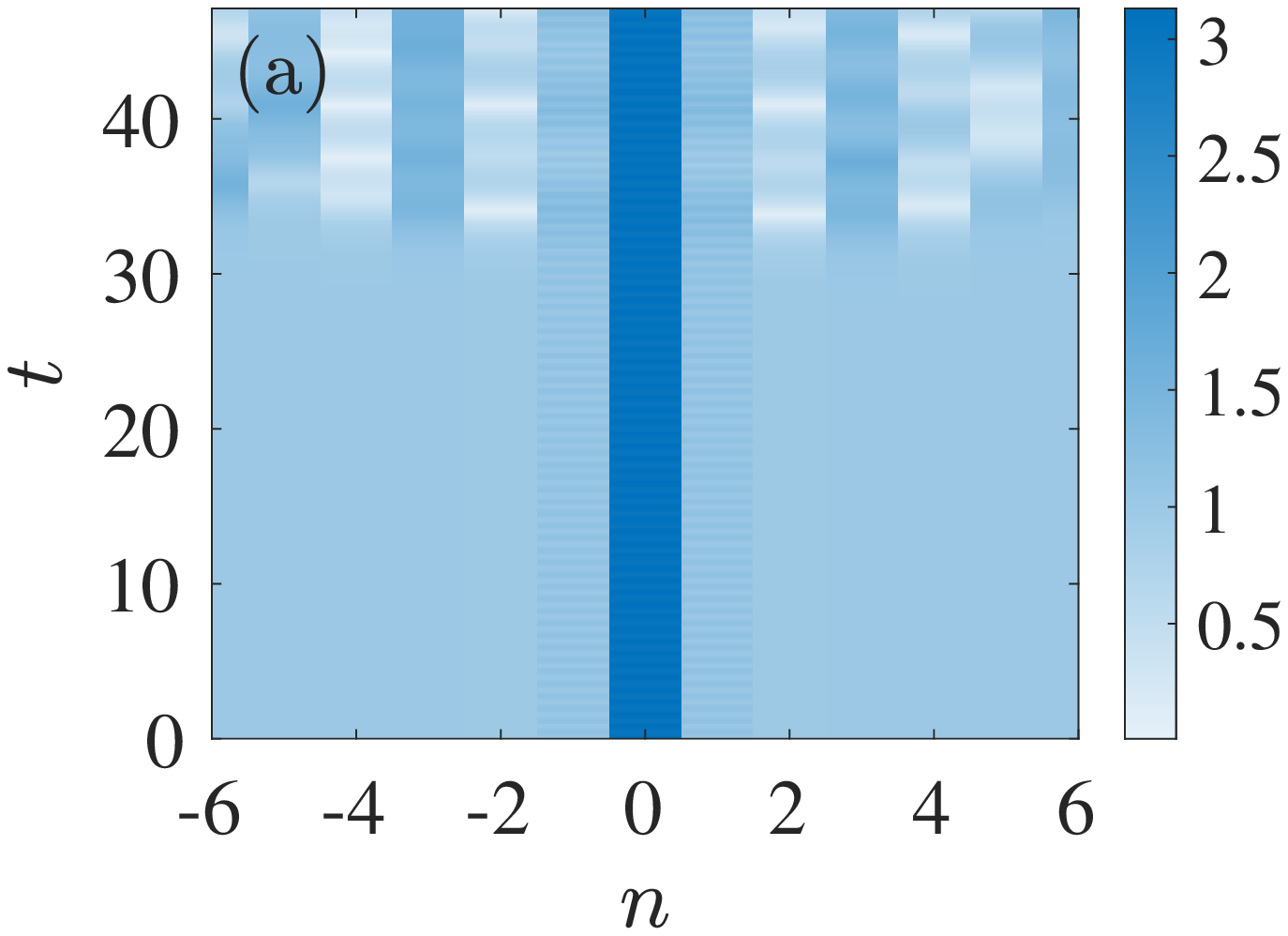} 
\includegraphics[height=.135\textheight, angle =0]{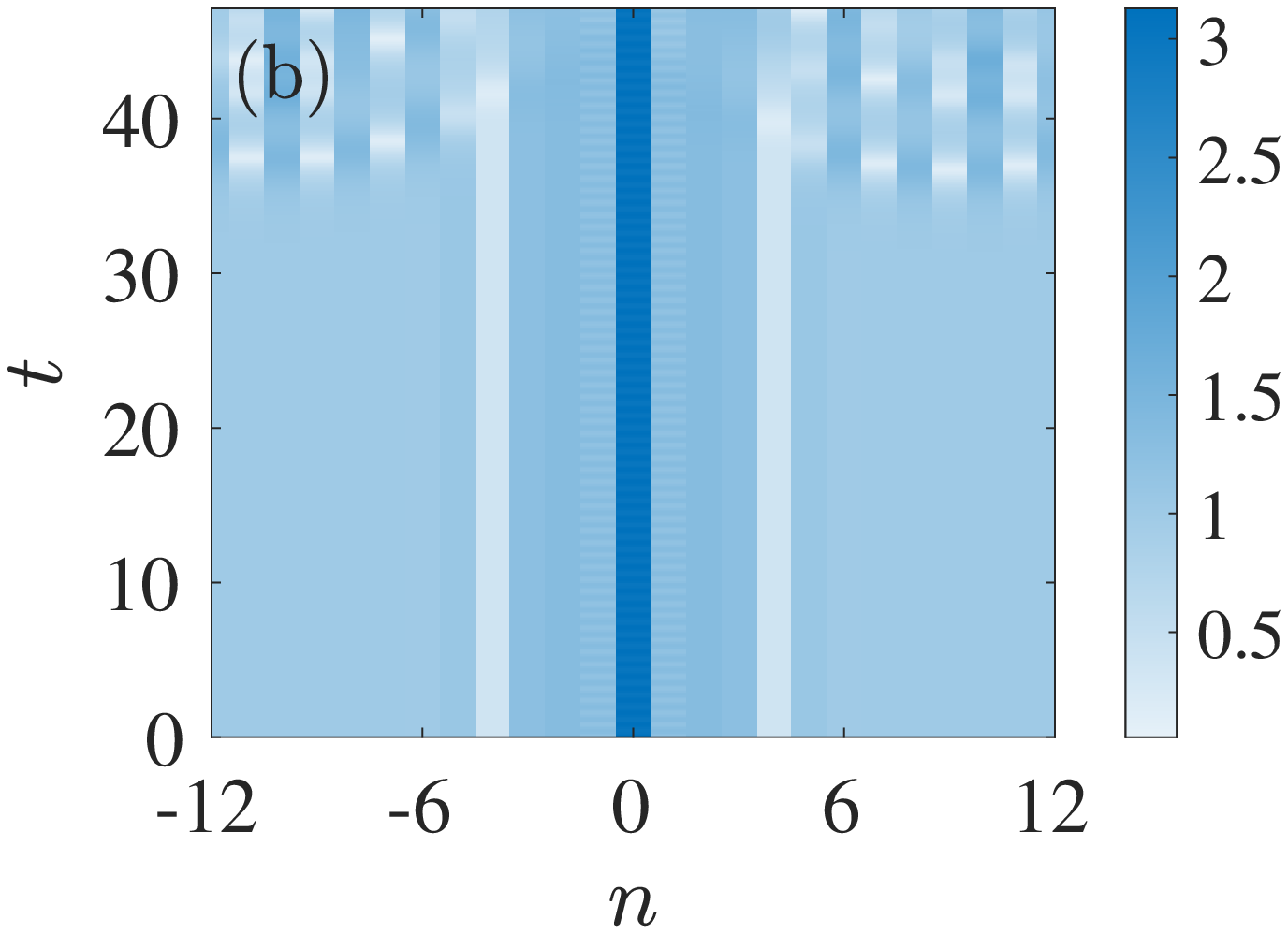} 
\includegraphics[height=.135\textheight, angle =0]{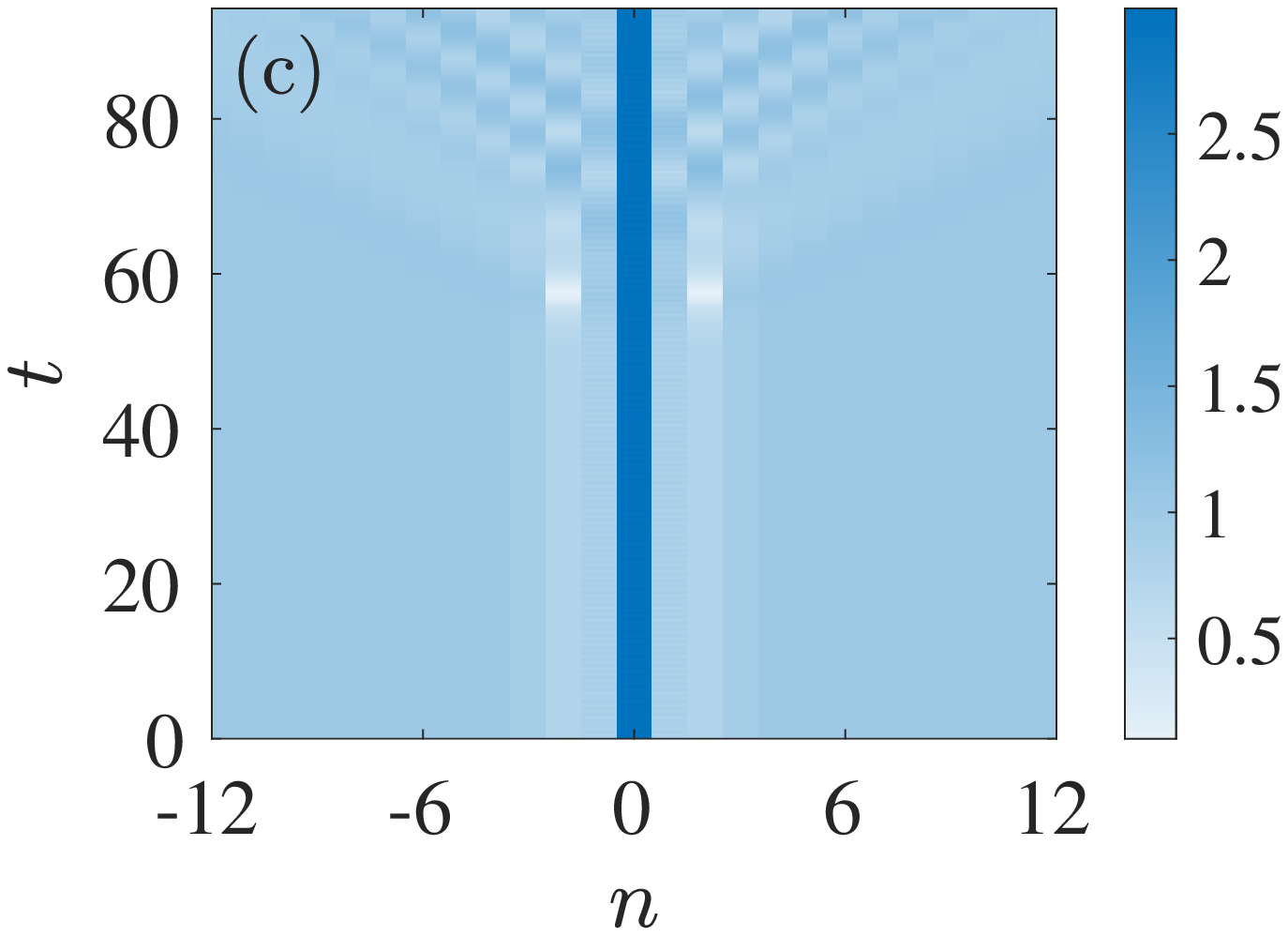} 
\includegraphics[height=.135\textheight, angle =0]{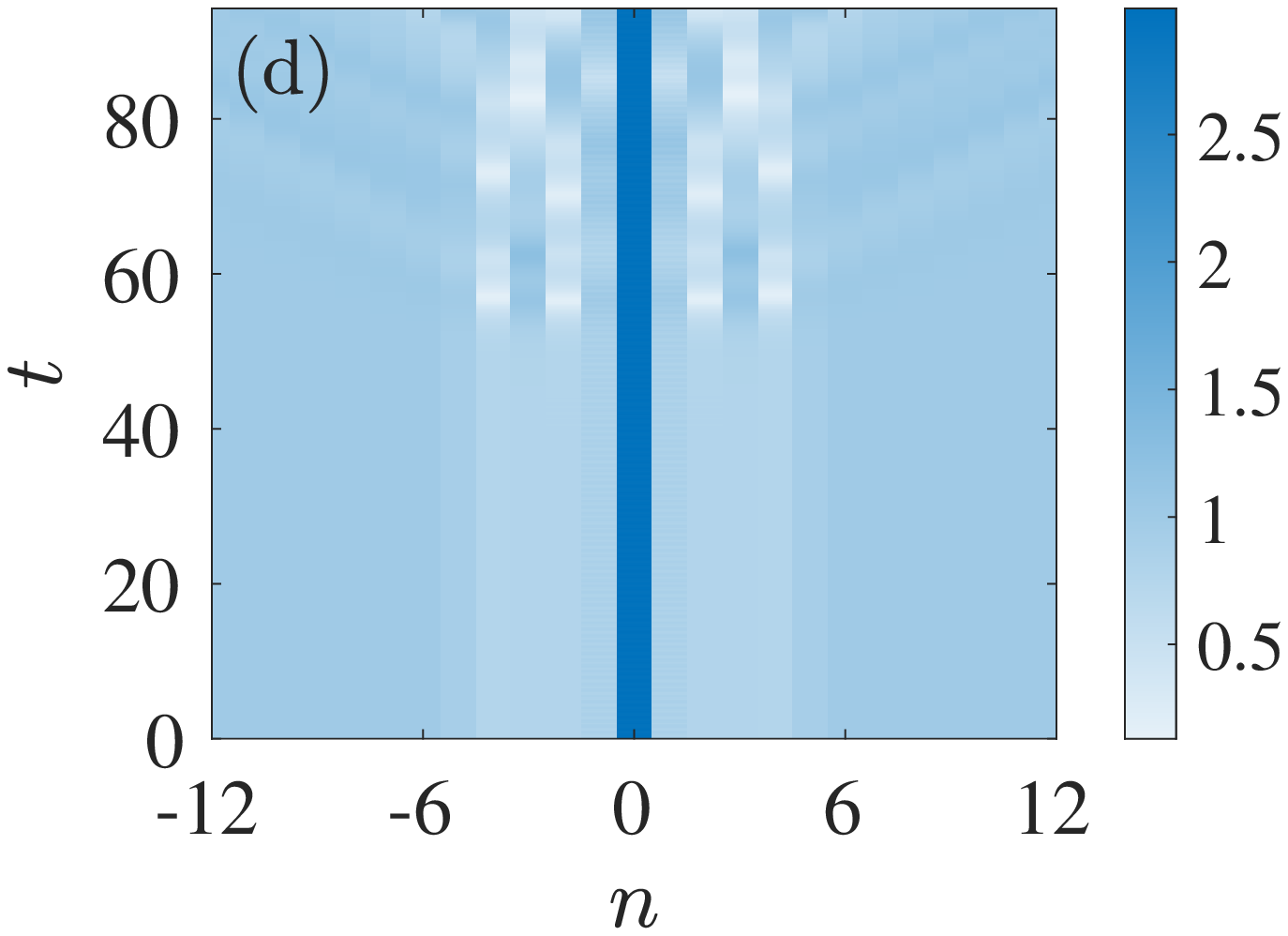} 
\end{center}
\caption{(Color online)
Spatio-temporal evolution of the amplitude $|u_{n}|$ of the quasiperiodic
solutions to the DNLS associated with the top panel of Fig.~\ref{fig1} with
$\omega_b=8$ (see also Fig.~\ref{fig5}). The labels (a)-(d) herein connect with
the solutions of Fig.~\ref{fig1} labeled with (a) ($d=0.2$), (d) ($d=0.2$),
(f) ($d=-0.1$), and (h) ($d=-0.1$), respectively.
}
\label{fig4}
\end{figure}

\begin{figure}[!pt]
\begin{center}
\includegraphics[height=.125\textheight, angle =0]{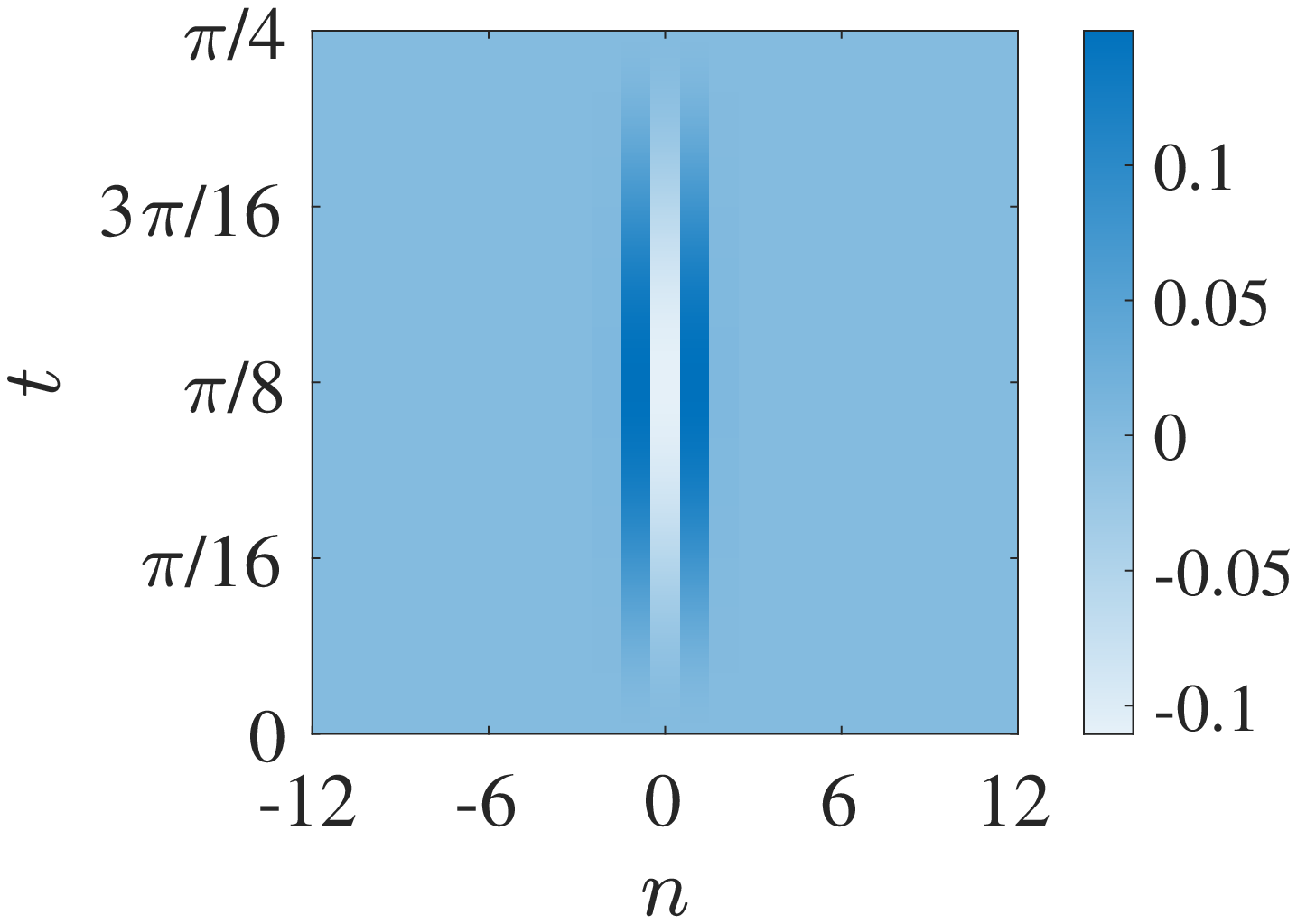}
\includegraphics[height=.125\textheight, angle =0]{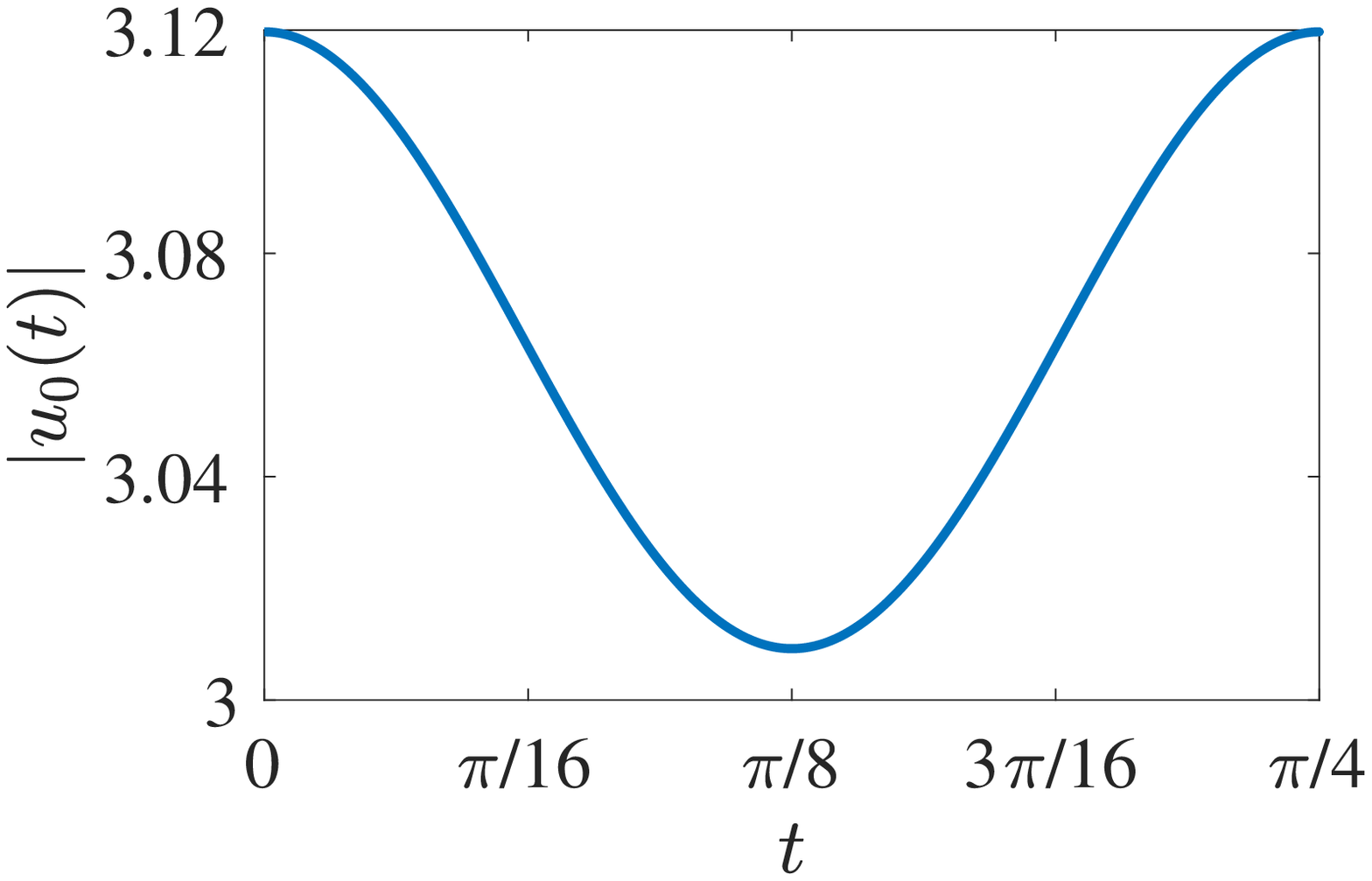}
\includegraphics[height=.125\textheight, angle =0]{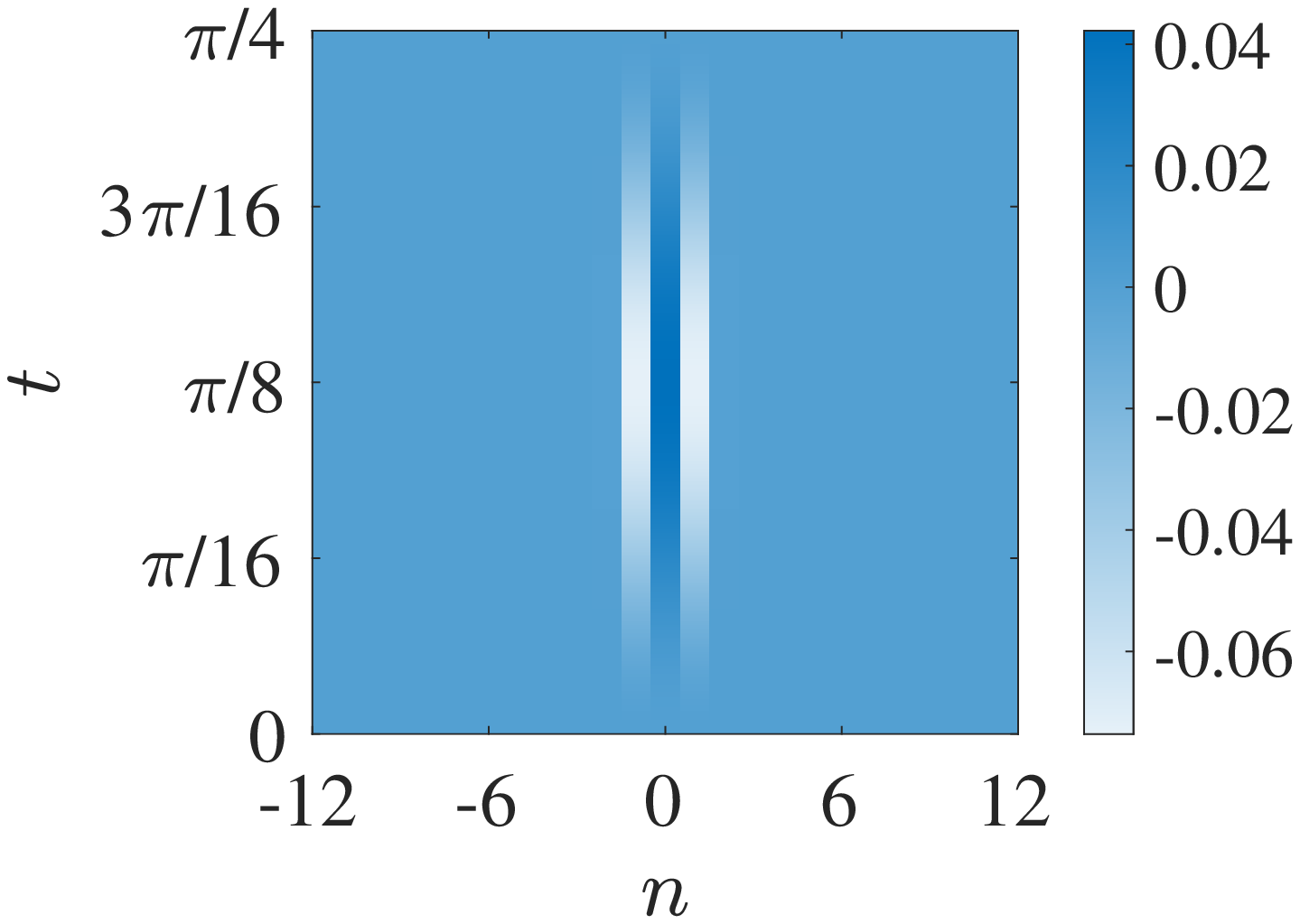}
\includegraphics[height=.125\textheight, angle =0]{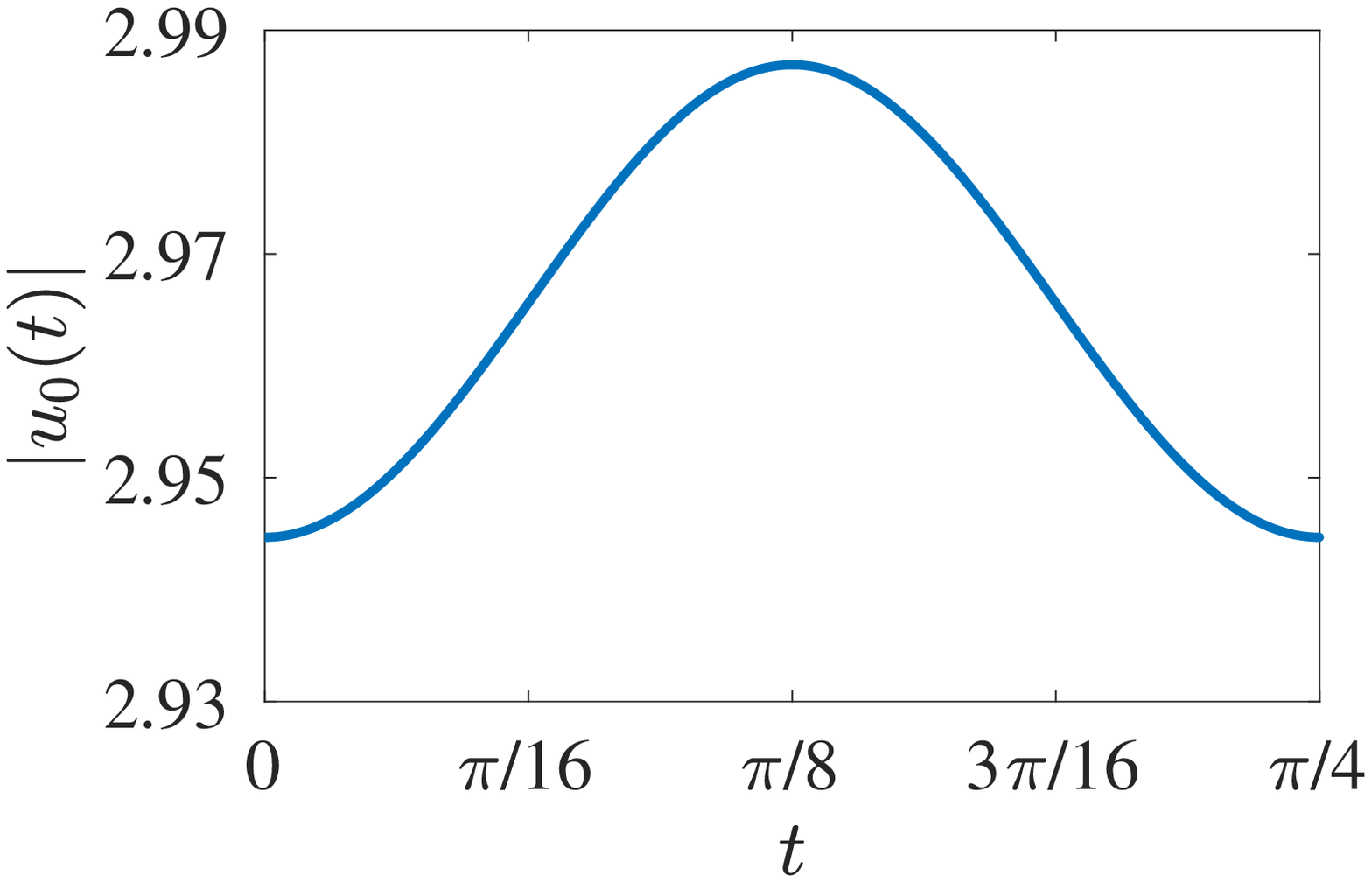}
\end{center}
\caption{(Color online) Complementary numerical results associated with the
dynamics presented in Fig.~\ref{fig4}. The left and right columns present
the difference of the amplitude $|u_{n}(t)|-|u_{n}(t=0)|$ and temporal
evolution of $|u_{0}(t)|$, i.e., the amplitude of the solution at the
center of the lattice, respectively. The top and bottom rows of the figure
correspond to the time-periodic solutions associated with the states (a) and
(f) shown in the top panel of Fig.~\ref{fig1} (whose spatio-temporal evolution is
presented in Figs.~\ref{fig4}(a) and (d), respectively).
}
\label{fig5}
\end{figure}

Finally, we move to the results on the spatio-temporal evolution of
the quasiperiodic (in the original frame --- time-periodic in their
modulus evolution) solutions presented in Figs.~\ref{fig4} and~\ref{fig5}
(again, with $\omega_{b}=8$ and $\Omega=1$) both for the focusing and defocusing cases.
We only considered site-centered
breathers since the results for bond-centered ones are similar and are omitted herein.
The panels (a)-(d) in Fig.~\ref{fig4} showcase the spatio-temporal evolution of the
amplitude $|u_{n}|$ of (site-centered) breathers which respectively connect with the
labels (a) and (d) (focusing regime) as well as (f) and (h) (defocusing regime) of the
top panel of Fig.~\ref{fig1}. We note in passing that we depict the amplitude and not the
density, i.e., $|u_{n}|^{2}$ of the solutions therein due to the dim variation of the profiles
over a period.
We used the breather our Newton solver converged to as an
initial condition, and integrated Eq.~\eqref{dnls_numer_1} forward in time.
The terminal times for
the results shown in panels (a), (b), and (d) are $60\,T$, and $120\,T$ for (c).

It can be discerned from these panels that all solutions are dynamically unstable although
the ones shown in panels (a) and (b) (being examples of the focusing regime) are modulationally
unstable per our Floquet stability analysis (see, their Floquet spectra in (a) and (f) in the
top panel of Fig.~\ref{fig1}). Around $t\approx 33$ and $t\approx 35$ in Figs.~\ref{fig4}(a)
and (b), the instability of the background (due to MI) manifests
itself and appears
to also partially affect
the core structure of the solution. The center of the solution, i.e.,
at $n=0$ remains
chiefly
unaltered even up to $t=1000\,T$ although the remaining
nodes that were originally deviated
from the background are more drastically modified
(results not shown). On the contrary, in Figs.~\ref{fig4}(c)
and (d), i.e., defocusing regime, the instability is emanating now from the core structure
of the solution (i.e., from the point spectrum instability of the
state,
while the background is in this case modulationally stable), see the
Floquet spectra in (f) and (h) in the top panel of
Fig.~\ref{fig1}. Interestingly,
however, and somewhat similarly with the
panels (a) and (b) in the figure, the solutions at $n=0$ in panels (c)
and (d) remain
robust and mostly unaltered in this defocusing case too.
These defocusing regime
instabilities manifest themselves at later times. This is due to the
fact that fewer
unstable
eigendirections emerge in the defocusing regime, and simultaneously their growth rates are smaller
compared to the ones in the focusing one. Indicatively, in Figs.~\ref{fig4}(c) and (d), the dominant
unstable eigendirection respectively corresponds to $\lambda_{r}\approx 1.517$ and $\lambda_{r}\approx 1.544$
as compared with $\lambda_{r}\approx 2.157$ of Figs.~\ref{fig4}(a) and (b).

We complement the results of Fig.~\ref{fig4} with the panels of Fig.~\ref{fig5}. The left
and right columns of the figure demonstrate the spatio-temporal dependence of the difference
of the amplitudes $|u_{n}(t)|-|u_{n}(t=0)|$ and temporal evolution of the amplitude of the central
site, i.e., $|u_{n}(t)|$, respectively, for the cases of Figs.~\ref{fig4}(a) (top row, focusing regime)
and (b) (bottom row, defocusing regime). Recall that in this work, we were motivated by the possibility
of identifying a localized in-space, time-quasiperiodic (i.e.,
periodic in its modulus) solution that would share some of the
features
of rogue waves (solutions of
extreme amplitude that ``appear out of nowhere and disappear without a trace'') of the KM type that sits
on a {flat} background.
Our original motivation  involved waveforms that would share these
characteristics,
designed in an ``on-demand'' way (as similar to a KM or Peregrine
waveform) at the AC limit. The dynamics of Fig.~\ref{fig5}, however,
demonstrate that the amplitude of the oscillations of the solutions
does not share the ``extreme'' feature of the rogue patterns but
is, instead,  rather ``small'' (compared to the
size of the background), thus distinguishing between the
patterns identified and the continuum (as well as integrable discrete)
rogue ones.
Nevertheless, the waveforms identified are novel quasiperiodic ones on a
finite
background that should be accessible, in principle, in optical or
atomic
settings where the DNLS is the suitable physical model.
Importantly, also, and while the AC limit may not provide as
straightforward
of a path for an on-demand construction of rogue patterns,
it still remains an open question whether
KM breathers and Peregrine solitons on a non-vanishing background
could be identified for the physically relevant DNLS model at {\it
  finite coupling}. We discuss this central open question further in the
concluding section that follows.

\section{Conclusions \& Future Challenges}
In the present work we have explored the interface between two exciting
recent directions, namely the study of DNLS models with an eye to applications
in optical and atomic physics, and the potential of formation of rogue
wave-like structures in dispersive nonlinear systems, utilizing the firm
analytical handle on the latter provided by the anti-continuum limit of uncoupled
lattice sites. We used a rigorous argument to showcase that relevant spatially
localized but temporally quasiperiodic (periodic in the modulus) solutions
should exist in such models. We illustrated a surprising continuity between
the branches of associated solutions in the modulationally unstable self-focusing
and the modulationally stable self-defocusing nonlinearities. This led to a
snake-like bifurcation diagram featuring turning points between pairs of branches
for each sign of the coupling. The defocusing solution branches were found to
feature isolated (and potentially weak) instability, which, in turn, might
facilitate the emergence of such states in experiments. We also examined the
continuation of the relevant waveforms in the breather frequency (motivated by
the corresponding continuation of the KM breather towards the Peregrine soliton),
finding that while such a path can be meaningful in the focusing case, it always
leads to a turning point and the absence of such solutions near $\omega_b \rightarrow 0$
for the self-defocusing realm. Finally, the dynamics of such solutions were illustrated
to make the point of their controllable (according to their respective instability
growth rate) persistence, most notably in the modulationally stable self-defocusing
realm that may enable their potential future experimental observation. Nevertheless,
as explained in detail, while our solutions are interesting novel quasiperiodic patterns
on a finite background within the DNLS model, they do not share the full  spectrum of
rogue wave features and can thus not be characterized as ones such.

Naturally, this study raises a wide variety of additional questions
and forges potential avenues for further explorations. On the one hand, it
would be particularly interesting to explore the departure of the different
(integrable and non-integrable) DNLS models from the continuum limit.
Admittedly, this requires a different set of tools than the AC ones
leveraged herein, yet it is an important question whose answer may
shed light on the possibility of emergence of KM solutions to the (non-integrable)
DNLS  and even regarding the feasibility of their limiting
Peregrine profile within the DNLS model. On the integrable systems realm,
another relevant point concerns the relatively recent observation~\cite{Ohta_2014}
regarding the exact analytical rogue waves of the {\it defocusing}
Ablowitz-Ladik model. As these authors point out, the existence of
rogue waves in the latter setting is surprising (and such waves may
have also unexpected features such as a potential blowup in finite time).
The potential extension of such waves in the context of the Salerno
and eventually the DNLS model (in comparison with the waveforms considered
herein) would also be relevant to consider. Furthermore, an interesting feature
of the present considerations is that they are not dimensionally-dependent
(contrary to integrability-related considerations), hence the breathers
identified herein should persist to higher-dimensional cases and their
properties and dynamics therein constitute another relevant vein of
research. Similarly, whether quasiperiodic solutions on a finite
background can be spectrally stable is yet another question of
future interest. Such topics are currently under consideration and will
be reported in future works.

\begin{acknowledgments}
This work has been supported by the U.S. National Science
Foundation under Grants No. DMS-2204782 (E.G.C.), and
DMS-2110030 and DMS-2204702 (P.G.K.). J.C.-M.~acknowledges support from the EU (FEDER program 2014-2020) through both Consejería de Economía, Conocimiento, Empresas y Universidad de la Junta de Andalucía (under the project US-1380977), and MCIN/AEI/10.13039/501100011033 (under the projects PID2019-110430GB-C21 and PID2020-112620GB-I00). EGC expresses his
gratitude to D.~Pelinovsky (McMaster University) and A.~Scheel
(University of Minnesota) for fruitful discussions.
\end{acknowledgments}

\bibliographystyle{apsrev4-1}
\bibliography{main.bib}

\begin{thebibliography}{41}%
\makeatletter
\providecommand \@ifxundefined [1]{%
 \@ifx{#1\undefined}
}%
\providecommand \@ifnum [1]{%
 \ifnum #1\expandafter \@firstoftwo
 \else \expandafter \@secondoftwo
 \fi
}%
\providecommand \@ifx [1]{%
 \ifx #1\expandafter \@firstoftwo
 \else \expandafter \@secondoftwo
 \fi
}%
\providecommand \natexlab [1]{#1}%
\providecommand \enquote  [1]{``#1''}%
\providecommand \bibnamefont  [1]{#1}%
\providecommand \bibfnamefont [1]{#1}%
\providecommand \citenamefont [1]{#1}%
\providecommand \href@noop [0]{\@secondoftwo}%
\providecommand \href [0]{\begingroup \@sanitize@url \@href}%
\providecommand \@href[1]{\@@startlink{#1}\@@href}%
\providecommand \@@href[1]{\endgroup#1\@@endlink}%
\providecommand \@sanitize@url [0]{\catcode `\\12\catcode `\$12\catcode
  `\&12\catcode `\#12\catcode `\^12\catcode `\_12\catcode `\%12\relax}%
\providecommand \@@startlink[1]{}%
\providecommand \@@endlink[0]{}%
\providecommand \url  [0]{\begingroup\@sanitize@url \@url }%
\providecommand \@url [1]{\endgroup\@href {#1}{\urlprefix }}%
\providecommand \urlprefix  [0]{URL }%
\providecommand \Eprint [0]{\href }%
\providecommand \doibase [0]{http://dx.doi.org/}%
\providecommand \selectlanguage [0]{\@gobble}%
\providecommand \bibinfo  [0]{\@secondoftwo}%
\providecommand \bibfield  [0]{\@secondoftwo}%
\providecommand \translation [1]{[#1]}%
\providecommand \BibitemOpen [0]{}%
\providecommand \bibitemStop [0]{}%
\providecommand \bibitemNoStop [0]{.\EOS\space}%
\providecommand \EOS [0]{\spacefactor3000\relax}%
\providecommand \BibitemShut  [1]{\csname bibitem#1\endcsname}%
\let\auto@bib@innerbib\@empty
\bibitem [{\citenamefont {Sievers}\ and\ \citenamefont
  {Takeno}(1988)}]{takeno}%
  \BibitemOpen
  \bibfield  {author} {\bibinfo {author} {\bibfnamefont {A.~J.}\ \bibnamefont
  {Sievers}}\ and\ \bibinfo {author} {\bibfnamefont {S.}~\bibnamefont
  {Takeno}},\ }\href {\doibase 10.1103/PhysRevLett.61.970} {\bibfield
  {journal} {\bibinfo  {journal} {Phys. Rev. Lett.}\ }\textbf {\bibinfo
  {volume} {61}},\ \bibinfo {pages} {970} (\bibinfo {year} {1988})}\BibitemShut
  {NoStop}%
\bibitem [{\citenamefont {MacKay}\ and\ \citenamefont {Aubry}(1994)}]{mackay}%
  \BibitemOpen
  \bibfield  {author} {\bibinfo {author} {\bibfnamefont {R.~S.}\ \bibnamefont
  {MacKay}}\ and\ \bibinfo {author} {\bibfnamefont {S.}~\bibnamefont {Aubry}},\
  }\href {\doibase 10.1088/0951-7715/7/6/006} {\bibfield  {journal} {\bibinfo
  {journal} {Nonlinearity}\ }\textbf {\bibinfo {volume} {7}},\ \bibinfo {pages}
  {1623} (\bibinfo {year} {1994})}\BibitemShut {NoStop}%
\bibitem [{\citenamefont {Flach}\ and\ \citenamefont
  {Gorbach}(2008)}]{Flach2008}%
  \BibitemOpen
  \bibfield  {author} {\bibinfo {author} {\bibfnamefont {S.}~\bibnamefont
  {Flach}}\ and\ \bibinfo {author} {\bibfnamefont {A.~V.}\ \bibnamefont
  {Gorbach}},\ }\href {\doibase 10.1016/j.physrep.2008.05.002} {\bibfield
  {journal} {\bibinfo  {journal} {Physics Reports}\ }\textbf {\bibinfo {volume}
  {467}},\ \bibinfo {pages} {1} (\bibinfo {year} {2008})}\BibitemShut {NoStop}%
\bibitem [{\citenamefont {Aubry}(2006)}]{Aubry2006}%
  \BibitemOpen
  \bibfield  {author} {\bibinfo {author} {\bibfnamefont {S.}~\bibnamefont
  {Aubry}},\ }\href {\doibase 10.1016/j.physd.2005.12.020} {\bibfield
  {journal} {\bibinfo  {journal} {Physica D}\ }\textbf {\bibinfo {volume}
  {216}},\ \bibinfo {pages} {1} (\bibinfo {year} {2006})}\BibitemShut {NoStop}%
\bibitem [{\citenamefont {Lederer}\ \emph {et~al.}(2008)\citenamefont
  {Lederer}, \citenamefont {Stegeman}, \citenamefont {Christodoulides},
  \citenamefont {Assanto}, \citenamefont {Segev},\ and\ \citenamefont
  {Silberberg}}]{LEDERER20081}%
  \BibitemOpen
  \bibfield  {author} {\bibinfo {author} {\bibfnamefont {F.}~\bibnamefont
  {Lederer}}, \bibinfo {author} {\bibfnamefont {G.~I.}\ \bibnamefont
  {Stegeman}}, \bibinfo {author} {\bibfnamefont {D.~N.}\ \bibnamefont
  {Christodoulides}}, \bibinfo {author} {\bibfnamefont {G.}~\bibnamefont
  {Assanto}}, \bibinfo {author} {\bibfnamefont {M.}~\bibnamefont {Segev}}, \
  and\ \bibinfo {author} {\bibfnamefont {Y.}~\bibnamefont {Silberberg}},\
  }\href {\doibase https://doi.org/10.1016/j.physrep.2008.04.004} {\bibfield
  {journal} {\bibinfo  {journal} {Physics Reports}\ }\textbf {\bibinfo {volume}
  {463}},\ \bibinfo {pages} {1} (\bibinfo {year} {2008})}\BibitemShut {NoStop}%
\bibitem [{\citenamefont {Morsch}\ and\ \citenamefont
  {Oberthaler}(2006)}]{RevModPhys.78.179}%
  \BibitemOpen
  \bibfield  {author} {\bibinfo {author} {\bibfnamefont {O.}~\bibnamefont
  {Morsch}}\ and\ \bibinfo {author} {\bibfnamefont {M.}~\bibnamefont
  {Oberthaler}},\ }\href {\doibase 10.1103/RevModPhys.78.179} {\bibfield
  {journal} {\bibinfo  {journal} {Rev. Mod. Phys.}\ }\textbf {\bibinfo {volume}
  {78}},\ \bibinfo {pages} {179} (\bibinfo {year} {2006})}\BibitemShut
  {NoStop}%
\bibitem [{\citenamefont {Kevrekidis}(2009)}]{kev09}%
  \BibitemOpen
  \bibfield  {author} {\bibinfo {author} {\bibfnamefont {P.}~\bibnamefont
  {Kevrekidis}},\ }\href
  {https://link.springer.com/book/10.1007/978-3-540-89199-4} {\emph {\bibinfo
  {title} {{The discrete nonlinear Schr{\"o}dinger Equation}}}},\ \bibinfo
  {edition} {1st}\ ed.\ (\bibinfo  {publisher} {Springer-Verlag},\ \bibinfo
  {address} {Heidelberg},\ \bibinfo {year} {2009})\BibitemShut {NoStop}%
\bibitem [{\citenamefont {Kharif}\ \emph {et~al.}(2009)\citenamefont {Kharif},
  \citenamefont {Pelinovsky},\ and\ \citenamefont {Slunyaev}}]{Kharif2009}%
  \BibitemOpen
  \bibfield  {author} {\bibinfo {author} {\bibfnamefont {C.}~\bibnamefont
  {Kharif}}, \bibinfo {author} {\bibfnamefont {E.}~\bibnamefont {Pelinovsky}},
  \ and\ \bibinfo {author} {\bibfnamefont {A.}~\bibnamefont {Slunyaev}},\
  }\href {\doibase 10.1007/978-3-540-88419-4} {\emph {\bibinfo {title} {Rogue
  {{Waves}} in the {{Ocean}}}}},\ Advances in {{Geophysical}} and
  {{Environmental Mechanics}} and {{Mathematics}}\ (\bibinfo  {publisher}
  {{Springer-Verlag}},\ \bibinfo {address} {{Berlin Heidelberg}},\ \bibinfo
  {year} {2009})\BibitemShut {NoStop}%
\bibitem [{\citenamefont {Draper}(1966)}]{Draper1966}%
  \BibitemOpen
  \bibfield  {author} {\bibinfo {author} {\bibfnamefont {L.}~\bibnamefont
  {Draper}},\ }\href {\doibase 10.1002/j.1477-8696.1966.tb05176.x} {\bibfield
  {journal} {\bibinfo  {journal} {Weather}\ }\textbf {\bibinfo {volume} {21}},\
  \bibinfo {pages} {2} (\bibinfo {year} {1966})}\BibitemShut {NoStop}%
\bibitem [{\citenamefont {Solli}\ \emph {et~al.}(2007)\citenamefont {Solli},
  \citenamefont {Ropers}, \citenamefont {Koonath},\ and\ \citenamefont
  {Jalali}}]{Solli2007}%
  \BibitemOpen
  \bibfield  {author} {\bibinfo {author} {\bibfnamefont {D.~R.}\ \bibnamefont
  {Solli}}, \bibinfo {author} {\bibfnamefont {C.}~\bibnamefont {Ropers}},
  \bibinfo {author} {\bibfnamefont {P.}~\bibnamefont {Koonath}}, \ and\
  \bibinfo {author} {\bibfnamefont {B.}~\bibnamefont {Jalali}},\ }\href
  {\doibase 10.1038/nature06402} {\bibfield  {journal} {\bibinfo  {journal}
  {Nature}\ }\textbf {\bibinfo {volume} {450}},\ \bibinfo {pages} {1054}
  (\bibinfo {year} {2007})}\BibitemShut {NoStop}%
\bibitem [{\citenamefont {Solli}\ \emph {et~al.}(2008)\citenamefont {Solli},
  \citenamefont {Ropers},\ and\ \citenamefont {Jalali}}]{Solli2008}%
  \BibitemOpen
  \bibfield  {author} {\bibinfo {author} {\bibfnamefont {D.~R.}\ \bibnamefont
  {Solli}}, \bibinfo {author} {\bibfnamefont {C.}~\bibnamefont {Ropers}}, \
  and\ \bibinfo {author} {\bibfnamefont {B.}~\bibnamefont {Jalali}},\ }\href
  {\doibase 10.1103/PhysRevLett.101.233902} {\bibfield  {journal} {\bibinfo
  {journal} {Phys. Rev. Lett.}\ }\textbf {\bibinfo {volume} {101}},\ \bibinfo
  {pages} {233902} (\bibinfo {year} {2008})}\BibitemShut {NoStop}%
\bibitem [{\citenamefont {Kibler}\ \emph {et~al.}(2010)\citenamefont {Kibler},
  \citenamefont {Fatome}, \citenamefont {Finot}, \citenamefont {Millot},
  \citenamefont {Dias}, \citenamefont {Genty}, \citenamefont {Akhmediev},\ and\
  \citenamefont {Dudley}}]{Kibler2010}%
  \BibitemOpen
  \bibfield  {author} {\bibinfo {author} {\bibfnamefont {B.}~\bibnamefont
  {Kibler}}, \bibinfo {author} {\bibfnamefont {J.}~\bibnamefont {Fatome}},
  \bibinfo {author} {\bibfnamefont {C.}~\bibnamefont {Finot}}, \bibinfo
  {author} {\bibfnamefont {G.}~\bibnamefont {Millot}}, \bibinfo {author}
  {\bibfnamefont {F.}~\bibnamefont {Dias}}, \bibinfo {author} {\bibfnamefont
  {G.}~\bibnamefont {Genty}}, \bibinfo {author} {\bibfnamefont
  {N.}~\bibnamefont {Akhmediev}}, \ and\ \bibinfo {author} {\bibfnamefont
  {J.~M.}\ \bibnamefont {Dudley}},\ }\href {\doibase 10.1038/nphys1740}
  {\bibfield  {journal} {\bibinfo  {journal} {Nature Phys}\ }\textbf {\bibinfo
  {volume} {6}},\ \bibinfo {pages} {790} (\bibinfo {year} {2010})}\BibitemShut
  {NoStop}%
\bibitem [{\citenamefont {Kibler}\ \emph {et~al.}(2012)\citenamefont {Kibler},
  \citenamefont {Fatome}, \citenamefont {Finot}, \citenamefont {Millot},
  \citenamefont {Genty}, \citenamefont {Wetzel}, \citenamefont {Akhmediev},
  \citenamefont {Dias},\ and\ \citenamefont {Dudley}}]{Kibler2012}%
  \BibitemOpen
  \bibfield  {author} {\bibinfo {author} {\bibfnamefont {B.}~\bibnamefont
  {Kibler}}, \bibinfo {author} {\bibfnamefont {J.}~\bibnamefont {Fatome}},
  \bibinfo {author} {\bibfnamefont {C.}~\bibnamefont {Finot}}, \bibinfo
  {author} {\bibfnamefont {G.}~\bibnamefont {Millot}}, \bibinfo {author}
  {\bibfnamefont {G.}~\bibnamefont {Genty}}, \bibinfo {author} {\bibfnamefont
  {B.}~\bibnamefont {Wetzel}}, \bibinfo {author} {\bibfnamefont
  {N.}~\bibnamefont {Akhmediev}}, \bibinfo {author} {\bibfnamefont
  {F.}~\bibnamefont {Dias}}, \ and\ \bibinfo {author} {\bibfnamefont {J.~M.}\
  \bibnamefont {Dudley}},\ }\href {\doibase 10.1038/srep00463} {\bibfield
  {journal} {\bibinfo  {journal} {Sci. Rep.}\ }\textbf {\bibinfo {volume}
  {2}},\ \bibinfo {pages} {463} (\bibinfo {year} {2012})}\BibitemShut {NoStop}%
\bibitem [{\citenamefont {DeVore}\ \emph {et~al.}(2013)\citenamefont {DeVore},
  \citenamefont {Solli}, \citenamefont {Borlaug}, \citenamefont {Ropers},\ and\
  \citenamefont {Jalali}}]{DeVore2013}%
  \BibitemOpen
  \bibfield  {author} {\bibinfo {author} {\bibfnamefont {P.~T.~S.}\
  \bibnamefont {DeVore}}, \bibinfo {author} {\bibfnamefont {D.~R.}\
  \bibnamefont {Solli}}, \bibinfo {author} {\bibfnamefont {D.}~\bibnamefont
  {Borlaug}}, \bibinfo {author} {\bibfnamefont {C.}~\bibnamefont {Ropers}}, \
  and\ \bibinfo {author} {\bibfnamefont {B.}~\bibnamefont {Jalali}},\ }\href
  {\doibase 10.1088/2040-8978/15/6/064001} {\bibfield  {journal} {\bibinfo
  {journal} {J. Opt.}\ }\textbf {\bibinfo {volume} {15}},\ \bibinfo {pages}
  {064001} (\bibinfo {year} {2013})}\BibitemShut {NoStop}%
\bibitem [{\citenamefont {Frisquet}\ \emph {et~al.}(2016)\citenamefont
  {Frisquet}, \citenamefont {Kibler}, \citenamefont {Morin}, \citenamefont
  {Baronio}, \citenamefont {Conforti}, \citenamefont {Millot},\ and\
  \citenamefont {Wabnitz}}]{Frisquet2016}%
  \BibitemOpen
  \bibfield  {author} {\bibinfo {author} {\bibfnamefont {B.}~\bibnamefont
  {Frisquet}}, \bibinfo {author} {\bibfnamefont {B.}~\bibnamefont {Kibler}},
  \bibinfo {author} {\bibfnamefont {P.}~\bibnamefont {Morin}}, \bibinfo
  {author} {\bibfnamefont {F.}~\bibnamefont {Baronio}}, \bibinfo {author}
  {\bibfnamefont {M.}~\bibnamefont {Conforti}}, \bibinfo {author}
  {\bibfnamefont {G.}~\bibnamefont {Millot}}, \ and\ \bibinfo {author}
  {\bibfnamefont {S.}~\bibnamefont {Wabnitz}},\ }\href {\doibase
  10.1038/srep20785} {\bibfield  {journal} {\bibinfo  {journal} {Sci Rep}\
  }\textbf {\bibinfo {volume} {6}},\ \bibinfo {pages} {20785} (\bibinfo {year}
  {2016})}\BibitemShut {NoStop}%
\bibitem [{\citenamefont {Chabchoub}\ \emph {et~al.}(2011)\citenamefont
  {Chabchoub}, \citenamefont {Hoffmann},\ and\ \citenamefont
  {Akhmediev}}]{Chabchoub2011}%
  \BibitemOpen
  \bibfield  {author} {\bibinfo {author} {\bibfnamefont {A.}~\bibnamefont
  {Chabchoub}}, \bibinfo {author} {\bibfnamefont {N.~P.}\ \bibnamefont
  {Hoffmann}}, \ and\ \bibinfo {author} {\bibfnamefont {N.}~\bibnamefont
  {Akhmediev}},\ }\href {\doibase 10.1103/PhysRevLett.106.204502} {\bibfield
  {journal} {\bibinfo  {journal} {Phys. Rev. Lett.}\ }\textbf {\bibinfo
  {volume} {106}},\ \bibinfo {pages} {204502} (\bibinfo {year}
  {2011})}\BibitemShut {NoStop}%
\bibitem [{\citenamefont {Chabchoub}\ \emph {et~al.}(2012)\citenamefont
  {Chabchoub}, \citenamefont {Hoffmann}, \citenamefont {Onorato},\ and\
  \citenamefont {Akhmediev}}]{Chabchoub2012}%
  \BibitemOpen
  \bibfield  {author} {\bibinfo {author} {\bibfnamefont {A.}~\bibnamefont
  {Chabchoub}}, \bibinfo {author} {\bibfnamefont {N.}~\bibnamefont {Hoffmann}},
  \bibinfo {author} {\bibfnamefont {M.}~\bibnamefont {Onorato}}, \ and\
  \bibinfo {author} {\bibfnamefont {N.}~\bibnamefont {Akhmediev}},\ }\href
  {\doibase 10.1103/PhysRevX.2.011015} {\bibfield  {journal} {\bibinfo
  {journal} {Phys. Rev. X}\ }\textbf {\bibinfo {volume} {2}},\ \bibinfo {pages}
  {011015} (\bibinfo {year} {2012})}\BibitemShut {NoStop}%
\bibitem [{\citenamefont {Chabchoub}\ and\ \citenamefont
  {Fink}(2014)}]{Chabchoub2014}%
  \BibitemOpen
  \bibfield  {author} {\bibinfo {author} {\bibfnamefont {A.}~\bibnamefont
  {Chabchoub}}\ and\ \bibinfo {author} {\bibfnamefont {M.}~\bibnamefont
  {Fink}},\ }\href {\doibase 10.1103/PhysRevLett.112.124101} {\bibfield
  {journal} {\bibinfo  {journal} {Phys. Rev. Lett.}\ }\textbf {\bibinfo
  {volume} {112}},\ \bibinfo {pages} {124101} (\bibinfo {year}
  {2014})}\BibitemShut {NoStop}%
\bibitem [{\citenamefont {McAllister}\ \emph {et~al.}(2019)\citenamefont
  {McAllister}, \citenamefont {Draycott}, \citenamefont {Adcock}, \citenamefont
  {Taylor},\ and\ \citenamefont {van~den Bremer}}]{ton_2019}%
  \BibitemOpen
  \bibfield  {author} {\bibinfo {author} {\bibfnamefont {M.~L.}\ \bibnamefont
  {McAllister}}, \bibinfo {author} {\bibfnamefont {S.}~\bibnamefont
  {Draycott}}, \bibinfo {author} {\bibfnamefont {T.~A.~A.}\ \bibnamefont
  {Adcock}}, \bibinfo {author} {\bibfnamefont {P.~H.}\ \bibnamefont {Taylor}},
  \ and\ \bibinfo {author} {\bibfnamefont {T.~S.}\ \bibnamefont {van~den
  Bremer}},\ }\href {\doibase 10.1017/jfm.2018.886} {\bibfield  {journal}
  {\bibinfo  {journal} {Journal of Fluid Mechanics}\ }\textbf {\bibinfo
  {volume} {860}},\ \bibinfo {pages} {767–786} (\bibinfo {year}
  {2019})}\BibitemShut {NoStop}%
\bibitem [{\citenamefont {Bailung}\ \emph {et~al.}(2011)\citenamefont
  {Bailung}, \citenamefont {Sharma},\ and\ \citenamefont
  {Nakamura}}]{Bailung2011}%
  \BibitemOpen
  \bibfield  {author} {\bibinfo {author} {\bibfnamefont {H.}~\bibnamefont
  {Bailung}}, \bibinfo {author} {\bibfnamefont {S.~K.}\ \bibnamefont {Sharma}},
  \ and\ \bibinfo {author} {\bibfnamefont {Y.}~\bibnamefont {Nakamura}},\
  }\href {\doibase 10.1103/PhysRevLett.107.255005} {\bibfield  {journal}
  {\bibinfo  {journal} {Phys. Rev. Lett.}\ }\textbf {\bibinfo {volume} {107}},\
  \bibinfo {pages} {255005} (\bibinfo {year} {2011})}\BibitemShut {NoStop}%
\bibitem [{\citenamefont {Romero-Ros}\ \emph {et~al.}(2023)\citenamefont
  {Romero-Ros}, \citenamefont {Katsimiga}, \citenamefont {Mistakidis},
  \citenamefont {Mossman}, \citenamefont {Biondini}, \citenamefont
  {Schmelcher}, \citenamefont {Engels},\ and\ \citenamefont
  {Kevrekidis}}]{romeroros2023experimental}%
  \BibitemOpen
  \bibfield  {author} {\bibinfo {author} {\bibfnamefont {A.}~\bibnamefont
  {Romero-Ros}}, \bibinfo {author} {\bibfnamefont {G.~C.}\ \bibnamefont
  {Katsimiga}}, \bibinfo {author} {\bibfnamefont {S.~I.}\ \bibnamefont
  {Mistakidis}}, \bibinfo {author} {\bibfnamefont {S.}~\bibnamefont {Mossman}},
  \bibinfo {author} {\bibfnamefont {G.}~\bibnamefont {Biondini}}, \bibinfo
  {author} {\bibfnamefont {P.}~\bibnamefont {Schmelcher}}, \bibinfo {author}
  {\bibfnamefont {P.}~\bibnamefont {Engels}}, \ and\ \bibinfo {author}
  {\bibfnamefont {P.~G.}\ \bibnamefont {Kevrekidis}},\ }\href@noop {} {\enquote
  {\bibinfo {title} {Experimental realization of the peregrine soliton in
  repulsive two-component bose-einstein condensates},}\ } (\bibinfo {year}
  {2023}),\ \Eprint {http://arxiv.org/abs/2304.05951} {arXiv:2304.05951
  [nlin.PS]} \BibitemShut {NoStop}%
\bibitem [{\citenamefont {Yan}(2012)}]{Yan2012a}%
  \BibitemOpen
  \bibfield  {author} {\bibinfo {author} {\bibfnamefont {Z.}~\bibnamefont
  {Yan}},\ }\href {\doibase 10.1088/1742-6596/400/1/012084} {\bibfield
  {journal} {\bibinfo  {journal} {J. Phys.: Conf. Ser.}\ }\textbf {\bibinfo
  {volume} {400}},\ \bibinfo {pages} {012084} (\bibinfo {year}
  {2012})}\BibitemShut {NoStop}%
\bibitem [{\citenamefont {Onorato}\ \emph {et~al.}(2013)\citenamefont
  {Onorato}, \citenamefont {Residori}, \citenamefont {Bortolozzo},
  \citenamefont {Montina},\ and\ \citenamefont {Arecchi}}]{Onorato2013}%
  \BibitemOpen
  \bibfield  {author} {\bibinfo {author} {\bibfnamefont {M.}~\bibnamefont
  {Onorato}}, \bibinfo {author} {\bibfnamefont {S.}~\bibnamefont {Residori}},
  \bibinfo {author} {\bibfnamefont {U.}~\bibnamefont {Bortolozzo}}, \bibinfo
  {author} {\bibfnamefont {A.}~\bibnamefont {Montina}}, \ and\ \bibinfo
  {author} {\bibfnamefont {F.~T.}\ \bibnamefont {Arecchi}},\ }\href {\doibase
  10.1016/j.physrep.2013.03.001} {\bibfield  {journal} {\bibinfo  {journal}
  {Phys. Rep.}\ }\textbf {\bibinfo {volume} {528}},\ \bibinfo {pages} {47}
  (\bibinfo {year} {2013})}\BibitemShut {NoStop}%
\bibitem [{\citenamefont {Dudley}\ \emph {et~al.}(2014)\citenamefont {Dudley},
  \citenamefont {Dias}, \citenamefont {Erkintalo},\ and\ \citenamefont
  {Genty}}]{Dudley2014}%
  \BibitemOpen
  \bibfield  {author} {\bibinfo {author} {\bibfnamefont {J.~M.}\ \bibnamefont
  {Dudley}}, \bibinfo {author} {\bibfnamefont {F.}~\bibnamefont {Dias}},
  \bibinfo {author} {\bibfnamefont {M.}~\bibnamefont {Erkintalo}}, \ and\
  \bibinfo {author} {\bibfnamefont {G.}~\bibnamefont {Genty}},\ }\href
  {\doibase 10.1038/nphoton.2014.220} {\bibfield  {journal} {\bibinfo
  {journal} {Nature Photon}\ }\textbf {\bibinfo {volume} {8}},\ \bibinfo
  {pages} {755} (\bibinfo {year} {2014})}\BibitemShut {NoStop}%
\bibitem [{\citenamefont {Mihalache}(2017)}]{Mihalache2017}%
  \BibitemOpen
  \bibfield  {author} {\bibinfo {author} {\bibfnamefont {D.}~\bibnamefont
  {Mihalache}},\ }\href@noop {} {\bibfield  {journal} {\bibinfo  {journal}
  {Rom. Rep. Phys}\ }\textbf {\bibinfo {volume} {69}},\ \bibinfo {pages} {28}
  (\bibinfo {year} {2017})}\BibitemShut {NoStop}%
\bibitem [{\citenamefont {Dudley}\ \emph {et~al.}(2019)\citenamefont {Dudley},
  \citenamefont {Genty}, \citenamefont {Mussot}, \citenamefont {Chabchoub},\
  and\ \citenamefont {Dias}}]{natrevphys}%
  \BibitemOpen
  \bibfield  {author} {\bibinfo {author} {\bibfnamefont {J.~M.}\ \bibnamefont
  {Dudley}}, \bibinfo {author} {\bibfnamefont {G.}~\bibnamefont {Genty}},
  \bibinfo {author} {\bibfnamefont {A.}~\bibnamefont {Mussot}}, \bibinfo
  {author} {\bibfnamefont {A.}~\bibnamefont {Chabchoub}}, \ and\ \bibinfo
  {author} {\bibfnamefont {F.}~\bibnamefont {Dias}},\ }\href {\doibase
  10.1038/s42254-019-0100-0} {\bibfield  {journal} {\bibinfo  {journal} {Nat.
  Rev. Phys.}\ }\textbf {\bibinfo {volume} {1}},\ \bibinfo {pages} {675}
  (\bibinfo {year} {2019})}\BibitemShut {NoStop}%
\bibitem [{\citenamefont {Peregrine}(1983)}]{Peregrine1983}%
  \BibitemOpen
  \bibfield  {author} {\bibinfo {author} {\bibfnamefont {D.~H.}\ \bibnamefont
  {Peregrine}},\ }\href {\doibase 10.1017/S0334270000003891} {\bibfield
  {journal} {\bibinfo  {journal} {ANZIAM J.}\ }\textbf {\bibinfo {volume}
  {25}},\ \bibinfo {pages} {16} (\bibinfo {year} {1983})}\BibitemShut {NoStop}%
\bibitem [{\citenamefont {Kuznetsov}(1977)}]{Kuznetsov1977}%
  \BibitemOpen
  \bibfield  {author} {\bibinfo {author} {\bibfnamefont {E.~A.}\ \bibnamefont
  {Kuznetsov}},\ }\href@noop {} {\bibfield  {journal} {\bibinfo  {journal}
  {Sov. Phys.-Dokl.}\ }\textbf {\bibinfo {volume} {236}},\ \bibinfo {pages}
  {575} (\bibinfo {year} {1977})}\BibitemShut {NoStop}%
\bibitem [{\citenamefont {Ma}(1979)}]{Ma1979}%
  \BibitemOpen
  \bibfield  {author} {\bibinfo {author} {\bibfnamefont {Y.-C.}\ \bibnamefont
  {Ma}},\ }\href {\doibase 10.1002/sapm197960143} {\bibfield  {journal}
  {\bibinfo  {journal} {Stud. Appl. Math.}\ }\textbf {\bibinfo {volume} {60}},\
  \bibinfo {pages} {43} (\bibinfo {year} {1979})}\BibitemShut {NoStop}%
\bibitem [{\citenamefont {Akhmediev}\ and\ \citenamefont
  {Korneev}(1986)}]{Akhmediev1986}%
  \BibitemOpen
  \bibfield  {author} {\bibinfo {author} {\bibfnamefont {N.~N.}\ \bibnamefont
  {Akhmediev}}\ and\ \bibinfo {author} {\bibfnamefont {V.~I.}\ \bibnamefont
  {Korneev}},\ }\href {\doibase 10.1007/BF01037866} {\bibfield  {journal}
  {\bibinfo  {journal} {Theor Math Phys}\ }\textbf {\bibinfo {volume} {69}},\
  \bibinfo {pages} {1089} (\bibinfo {year} {1986})}\BibitemShut {NoStop}%
\bibitem [{\citenamefont {Ankiewicz}\ \emph {et~al.}(2010)\citenamefont
  {Ankiewicz}, \citenamefont {Akhmediev},\ and\ \citenamefont
  {Soto-Crespo}}]{sotoc}%
  \BibitemOpen
  \bibfield  {author} {\bibinfo {author} {\bibfnamefont {A.}~\bibnamefont
  {Ankiewicz}}, \bibinfo {author} {\bibfnamefont {N.}~\bibnamefont
  {Akhmediev}}, \ and\ \bibinfo {author} {\bibfnamefont {J.~M.}\ \bibnamefont
  {Soto-Crespo}},\ }\href {\doibase 10.1103/PhysRevE.82.026602} {\bibfield
  {journal} {\bibinfo  {journal} {Phys. Rev. E}\ }\textbf {\bibinfo {volume}
  {82}},\ \bibinfo {pages} {026602} (\bibinfo {year} {2010})}\BibitemShut
  {NoStop}%
\bibitem [{\citenamefont {Salerno}(1992)}]{salerno1992quantum}%
  \BibitemOpen
  \bibfield  {author} {\bibinfo {author} {\bibfnamefont {M.}~\bibnamefont
  {Salerno}},\ }\href@noop {} {\bibfield  {journal} {\bibinfo  {journal}
  {Physical Review A}\ }\textbf {\bibinfo {volume} {46}},\ \bibinfo {pages}
  {6856} (\bibinfo {year} {1992})}\BibitemShut {NoStop}%
\bibitem [{\citenamefont {Hoffmann}\ \emph {et~al.}(2018)\citenamefont
  {Hoffmann}, \citenamefont {Charalampidis}, \citenamefont {Frantzeskakis},\
  and\ \citenamefont {Kevrekidis}}]{HOFFMANN20183064}%
  \BibitemOpen
  \bibfield  {author} {\bibinfo {author} {\bibfnamefont {C.}~\bibnamefont
  {Hoffmann}}, \bibinfo {author} {\bibfnamefont {E.}~\bibnamefont
  {Charalampidis}}, \bibinfo {author} {\bibfnamefont {D.}~\bibnamefont
  {Frantzeskakis}}, \ and\ \bibinfo {author} {\bibfnamefont {P.}~\bibnamefont
  {Kevrekidis}},\ }\href {\doibase
  https://doi.org/10.1016/j.physleta.2018.08.014} {\bibfield  {journal}
  {\bibinfo  {journal} {Physics Letters A}\ }\textbf {\bibinfo {volume}
  {382}},\ \bibinfo {pages} {3064} (\bibinfo {year} {2018})}\BibitemShut
  {NoStop}%
\bibitem [{\citenamefont {{Sullivan, J.}}\ \emph {et~al.}(2020)\citenamefont
  {{Sullivan, J.}}, \citenamefont {{Charalampidis, E. G.}}, \citenamefont
  {{Cuevas-Maraver, J.}}, \citenamefont {{Kevrekidis, P. G.}},\ and\
  \citenamefont {{Karachalios, N. I.}}}]{sullivan}%
  \BibitemOpen
  \bibfield  {author} {\bibinfo {author} {\bibnamefont {{Sullivan, J.}}},
  \bibinfo {author} {\bibnamefont {{Charalampidis, E. G.}}}, \bibinfo {author}
  {\bibnamefont {{Cuevas-Maraver, J.}}}, \bibinfo {author} {\bibnamefont
  {{Kevrekidis, P. G.}}}, \ and\ \bibinfo {author} {\bibnamefont {{Karachalios,
  N. I.}}},\ }\href {\doibase 10.1140/epjp/s13360-020-00596-1} {\bibfield
  {journal} {\bibinfo  {journal} {Eur. Phys. J. Plus}\ }\textbf {\bibinfo
  {volume} {135}},\ \bibinfo {pages} {607} (\bibinfo {year}
  {2020})}\BibitemShut {NoStop}%
\bibitem [{\citenamefont {Johansson}\ and\ \citenamefont
  {Aubry}(1997)}]{Johansson_1997}%
  \BibitemOpen
  \bibfield  {author} {\bibinfo {author} {\bibfnamefont {M.}~\bibnamefont
  {Johansson}}\ and\ \bibinfo {author} {\bibfnamefont {S.}~\bibnamefont
  {Aubry}},\ }\href {\doibase 10.1088/0951-7715/10/5/008} {\bibfield  {journal}
  {\bibinfo  {journal} {Nonlinearity}\ }\textbf {\bibinfo {volume} {10}},\
  \bibinfo {pages} {1151} (\bibinfo {year} {1997})}\BibitemShut {NoStop}%
\bibitem [{\citenamefont {Kevrekidis}\ and\ \citenamefont
  {Weinstein}(2003)}]{pgkmiw}%
  \BibitemOpen
  \bibfield  {author} {\bibinfo {author} {\bibfnamefont {P.}~\bibnamefont
  {Kevrekidis}}\ and\ \bibinfo {author} {\bibfnamefont {M.}~\bibnamefont
  {Weinstein}},\ }\href {\doibase
  https://doi.org/10.1016/S0378-4754(02)00185-4} {\bibfield  {journal}
  {\bibinfo  {journal} {Mathematics and Computers in Simulation}\ }\textbf
  {\bibinfo {volume} {62}},\ \bibinfo {pages} {65} (\bibinfo {year} {2003})},\
  \bibinfo {note} {nonlinear Waves: Computation and Theory II}\BibitemShut
  {NoStop}%
\bibitem [{\citenamefont {{Wikipedia contributors}}(2023)}]{enwiki:1154847066}%
  \BibitemOpen
  \bibfield  {author} {\bibinfo {author} {\bibnamefont {{Wikipedia
  contributors}}},\ }\href
  {https://en.wikipedia.org/w/index.php?title=Rogue_wave&oldid=1154847066}
  {\enquote {\bibinfo {title} {Rogue wave --- {Wikipedia}{,} the free
  encyclopedia},}\ } (\bibinfo {year} {2023}),\ \bibinfo {note} {[Online;
  accessed 31-May-2023]}\BibitemShut {NoStop}%
\bibitem [{\citenamefont {Kuznetsov}(2023)}]{kuznetsov_book_2023}%
  \BibitemOpen
  \bibfield  {author} {\bibinfo {author} {\bibfnamefont {Y.~A.}\ \bibnamefont
  {Kuznetsov}},\ }\href
  {https://link.springer.com/book/10.1007/978-3-031-22007-4} {\emph {\bibinfo
  {title} {Elements of {A}pplied {B}ifurcation {T}heory}}},\ Applied
  {M}athematical {S}ciences\ (\bibinfo  {publisher} {Springer-Verlag},\
  \bibinfo {address} {{New York}},\ \bibinfo {year} {2023})\BibitemShut
  {NoStop}%
\bibitem [{\citenamefont {Kivshar}\ and\ \citenamefont
  {Peyrard}(1992)}]{kivpey}%
  \BibitemOpen
  \bibfield  {author} {\bibinfo {author} {\bibfnamefont {Y.~S.}\ \bibnamefont
  {Kivshar}}\ and\ \bibinfo {author} {\bibfnamefont {M.}~\bibnamefont
  {Peyrard}},\ }\href {\doibase 10.1103/PhysRevA.46.3198} {\bibfield  {journal}
  {\bibinfo  {journal} {Phys. Rev. A}\ }\textbf {\bibinfo {volume} {46}},\
  \bibinfo {pages} {3198} (\bibinfo {year} {1992})}\BibitemShut {NoStop}%
\bibitem [{\citenamefont {Vakhitov}\ and\ \citenamefont
  {Kolokolov}(1973)}]{vakhitov}%
  \BibitemOpen
  \bibfield  {author} {\bibinfo {author} {\bibfnamefont {N.}~\bibnamefont
  {Vakhitov}}\ and\ \bibinfo {author} {\bibfnamefont {A.}~\bibnamefont
  {Kolokolov}},\ }\href@noop {} {\bibfield  {journal} {\bibinfo  {journal}
  {Radiophys. Quant. Electron.}\ }\textbf {\bibinfo {volume} {16}},\ \bibinfo
  {pages} {783} (\bibinfo {year} {1973})}\BibitemShut {NoStop}%
\bibitem [{\citenamefont {Ohta}\ and\ \citenamefont {Yang}(2014)}]{Ohta_2014}%
  \BibitemOpen
  \bibfield  {author} {\bibinfo {author} {\bibfnamefont {Y.}~\bibnamefont
  {Ohta}}\ and\ \bibinfo {author} {\bibfnamefont {J.}~\bibnamefont {Yang}},\
  }\href {\doibase 10.1088/1751-8113/47/25/255201} {\bibfield  {journal}
  {\bibinfo  {journal} {Journal of Physics A: Mathematical and Theoretical}\
  }\textbf {\bibinfo {volume} {47}},\ \bibinfo {pages} {255201} (\bibinfo
  {year} {2014})}\BibitemShut {NoStop}%
\end{thebibliography}%

\end{document}